\documentclass[prc,twocolumn,showpacs,superscriptaddress,preprintnumbers,amsmath,amssymb,nofootinbib]{revtex4-1}
\usepackage{bm}
\usepackage{dcolumn}
\usepackage{multirow}
\usepackage{ifpdf}
\usepackage{url}
\usepackage{comment}

\ifpdf
\usepackage{graphicx} 
\usepackage{hyperref}
\else 
\usepackage[dvipdfmx]{graphicx} 
\usepackage[dvipdfmx]{hyperref}
\usepackage[dvipdfmx]{color}
\fi
\usepackage{color} 
\usepackage{ulem}

\makeatletter
\let\MYcaption\@makecaption
\makeatother
\usepackage{subcaption}
\makeatletter
\let\@makecaption\MYcaption
\makeatother

\usepackage{slashed}
\usepackage{bm}
\usepackage{dcolumn}
\usepackage{multirow}
\usepackage{ifpdf}
\usepackage{url}
\usepackage{here}
\usepackage{graphicx}
\usepackage{indentfirst}
\usepackage{listings}

\usepackage{ifthen}
\usepackage{kpfonts}
\usepackage{tgtermes}
\usepackage{amsmath,amssymb}
\usepackage{fancyhdr}
\usepackage{titletoc}
\usepackage{xcolor}

\usepackage{mathtools}

\newcommand{\dndeta}[1][flat]
{
    \ifthenelse{\equal{#1}{flat}}{$\left< dN_{\mathrm{ch}}/d\eta\right>$}{}
    \ifthenelse{\equal{#1}{vertical}}{$\left< \cfrac{dN_{\mathrm{ch}}}{d\eta\right>}$}{}
}

\newcommand{\snn}[1][nucleus]
{
    \ifthenelse{\equal{#1}{nucleus}}{$\sqrt{s_{\mathrm{NN}}}$}{}
    \ifthenelse{\equal{#1}{proton}}{$\sqrt{s}$}{}
}

\hypersetup{
colorlinks=true,
linkcolor=blue,
citecolor=blue,
urlcolor=blue
}

\begin{document}

\title{
Constraint on initial conditions of one-dimensional expanding fluids from nonlinear causality
}
\author{Tau Hoshino}
\email{t-hoshino-1d3@eagle.sophia.ac.jp}
\affiliation{%
Department of Physics, Sophia University, Tokyo 102-8554, Japan
}

\author{Tetsufumi Hirano}
\email{hirano@sophia.ac.jp}
\affiliation{%
Department of Physics, Sophia University, Tokyo 102-8554, Japan
}

\date{\today}

\begin{abstract}

The initial conditions of one-dimensional expanding viscous fluids in relativistic heavy-ion collisions are scrutinized in terms of nonlinear causality of the relativistic hydrodynamic equations.
Conventionally, it is believed that the matter generated in relativistic heavy-ion collisions starts to behave as a fluid all at once at some initial time.
However, it is by no means trivial how soon after the first contact of two high-energy nuclei the fluid picture can be applied.
It is demonstrated that one-dimensional expanding viscous fluids violate the necessary and the sufficient conditions of nonlinear causality at large departures from local equilibrium.
We therefore quantify the inverse Reynolds number to justify the validity of the hydrodynamic description.
The initial conditions are strictly constrained not to violate the causality conditions during the time evolution.
With the help of the transverse energies per rapidity measured at the BNL Relativistic Heavy Ion Collider
(RHIC) and the CERN Large Hadron Collider (LHC), we obtain the minimum initial proper time and the maximum energy density allowed by nonlinear causality.
This analysis strongly suggests that the initial stage of relativistic heavy-ion collisions needs to be described by a non-equilibrium framework other than relativistic dissipative hydrodynamics.

\end{abstract}


\maketitle

\section{Introduction}
\label{sec:intro}

A vast body of experimental data at the Large Hadron Collider (LHC) in CERN and at Relativistic Heavy-Ion Collider (RHIC) in Brookhaven National Laboratory (BNL) together with theoretical analysis of these data by using sophisticated dynamical models have revealed rich and profound physics of the quark gluon plasma (QGP). 
The QGP is a deconfined nuclear matter which occupied the early Universe $\approx 10$ micro seconds after the big bang and its temperature reaches $\approx 10^{12}$ Kelvin.
Hence the physics of the QGP elucidates a state of the matter composed of fundamental degree of freedom at extremely high temperature \cite{Yagi:2005yb}.

To extract bulk and transport properties of the QGP, relativistic hydrodynamics has been playing a vital role as a framework to describe dynamics of local equilibrium states.
Since the QGP is created as a transient state of matter in relativistic heavy-ion collisions, it is of crucial importance to develop an integrated dynamical model based on relativistic hydrodynamics which describes multiple stages, in particular, before and after hydrodynamic evolution.
In the final state of high-energy nuclear collisions, the description of dynamics can be changed from macroscopic relativistic hydrodynamics to microscopic kinetic theory for dilute hadron gases. This became a standard prescription in dynamical modeling once the importance of the hadronic corona was recognized \cite{Hirano:2005wx}.
On the other hand, it has not been established so far how to describe the pre-hydrodynamic stage, namely how to initialize hydrodynamic fields in relativistic hydrodynamics.
The hydrodynamic equations as first order differential equations with respect to time demand initial conditions. 
Conventionally, one either parametrizes the initial hydrodynamic fields so that final momentum distributions provide a reasonable reproduction of experimental data or employs a framework to describe the early stages of dynamics until the hydrodynamic initial time. 
It is known that the initial conditions of hydrodynamic models cannot be uniquely determined. 
One of the most reliable ways is to constrain the initial conditions through Bayesian parameter estimation \cite{Bernhard:2016tnd,Bernhard:2019bmu,Auvinen:2020mpc,Nijs:2020ors,Nijs:2020roc,Parkkila:2021tqq,JETSCAPE:2020mzn,Heffernan:2023gye,Shen:2023awv}.
However, a lack of the knowledge to describe the dynamics prior to hydrodynamic evolution could bring a systematic uncertainty of posterior distributions of the parameters even by using Bayesian estimation.

In this paper, we scrutinize whether the assigned initial conditions in hydrodynamic modeling are justified from a viewpoint of nonlinear causality in the relativistic hydrodynamic equations. 
To demonstrate it, we employ one-dimensional expanding fluids, check whether the hydrodynamic description is valid in a relativistic sense, and constrain the region of initial conditions.

Quantum chromodynamics (QCD) is the first-principles description of the dynamics of quarks and gluons, and its formulation unquestionably respects the causality of a relativistic field theory.
However, it is by no means self-evident that relativistic hydrodynamics, as an effective theory in the long wavelength limit which can be derived from QCD, obeys the causality.

The second-order hydrodynamics including Muller--Israel--Stewart type constitutive equations \cite{Muller:1967zza,Israel:1976tn,Israel:1979wp} are known to obey causality at least within a linear regime as long as the relaxation time is sufficiently large.
This was shown for the first time in Ref.~\cite{Hiscock:1983zz} within propagation of a linear perturbation around a static  and uniform local equilibrium background in which energy density and pressure are constant ($e=$ const.~and $p=$ const.), flow velocity vanishes [$u^\mu = (1, 0, 0, 0)$], and the dissipative currents such as shear stress tensor and bulk pressure vanish ($\pi^{\mu \nu}=0$ and $\Pi = 0$).

However, the essential aspects of the second-order terms cannot be captured in the linear perturbation since these dissipative terms are higher order in small quantities and neglected in the linear regime.
Since the second-order hydrodynamic equations are a set of quasi-linear partial differential equation, causal propagation and local stability can be discussed without relying on small deviation.
Therefore it is of particular importance to go beyond the linear regime and to capture full nonlinearity of relativistic dissipative hydrodynamic equations.
To our best knowledge, the first study of nonlinear causality in the second-order relativistic hydrodynamic equations was done in one-dimensional expanding system in Ref.~\cite{HISCOCK1989125}.
Later some attempts were made to investigate a relation between causality and stability of the solutions of nonlinear hydrodynamic equations \cite{Denicol:2008ha,Pu:2009fj,Floerchinger:2017cii}.  
Recently, necessary and sufficient conditions were obtained from full nonlinear hydrodynamic equations in the nonlinear regime \cite{bemfica2021nonlinear}.
Since these conditions are written as inequalities including thermodynamic variables and dissipative currents, these can be used to (in-)validate the numerical solutions of hydrodynamics equations \cite{plumberg2022causality,Chiu:2021muk}.
These could provide new theoretical guidance on relativistic hydrodynamic simulations and robust Bayesian inference towards quantitative extraction of the properties of the QGP \cite{daSilva:2022xwu,ExTrEMe:2023nhy,Domingues:2024pom}. 

After a brief summary of the second-order hydrodynamics  in Sec.~\ref{sec:model}, we show the necessary and the sufficient conditions, constraints on initial conditions, and the equation-of-state dependence of one-dimensional expanding fluids in Sec.~\ref{sec:results}.
Section \ref{sec:summary} is devoted to a summary and conclusion of the present study. We also discuss more details about the one-dimensional expanding conformal fluids in Appendix \ref{sec:hydrodynamization} and review the necessary and the sufficient conditions of nonlinear causality in Appendix \ref{sec:causality appendix}.

Throughout this paper, we use natural units, $c=\hbar=k_B=1$, and Minkowski metric, $g_{\mu\nu} = \mathrm{diag}(1,-1,-1,-1)$.

\section{Model}
\label{sec:model}

Relativistic hydrodynamics conventionally describes space-time evolution of thermodynamic variables, dissipative currents, and fluid four-velocity under an assumption that the local system is close to thermal equilibrium. 
One can also regard its framework as an effective theory to describe long-time and long-range behaviors for a given microscopic system based on the conservation laws of energy and momentum and the gradient expansion. 
In the current framework of the second-order relativistic hydrodynamics, dissipative currents are promoted to dynamical variables.
Therefore one has to scrutinize whether given initial conditions, in particular, for dissipative currents are within a valid range of the framework or not. 

In this section, we review a framework of relativistic hydrodynamics in Sec.~\ref{sec:relativistic hydrodynamics} and introduce two equations of state employed in this study in Sec.~\ref{sec:EoS}.
Assuming translational and rotational symmetries in the transverse plane and boost invariance in the longitudinal direction, we obtain the equations of motion of the fluids in Sec.~\ref{sec:EoM}.
We demonstrate the behaviors of the solution of  in Sec.~\ref{sec:time_evolution_of_Reynolds_no}.

\subsection{Relativistic hydrodynamics}
\label{sec:relativistic hydrodynamics}

In the following, we assume the system does not contain any conserved charges and neglect the bulk pressure for simplicity. 
Hydrodynamic equations describe the continuity equations of energy and momentum conservation,   
\begin{align}
\label{eq:energy-momentum-conservation}
    \partial_\mu T^{\mu\nu}=0,\quad
        T^{\mu\nu} = e u^\mu u^\nu -p \Delta ^{\mu\nu}+\pi^{\mu\nu},
\end{align}
where $T^{\mu\nu}$ is the energy-momentum tensor,
$e$ is the energy density, $p$ is the pressure, $u^\mu$ is the fluid four-velocity, and  $\pi^{\mu\nu}$ is the shear stress tensor.
The fluid four-velocity is normalized as $u_\mu u^\mu = 1$.
By decomposing $T^{\mu \nu}$ into the time-like and the space-like components for each Lorentz index, these can be defined as
\begin{align}
    e \equiv u_\alpha T^{\alpha \beta}u_\beta,\quad 
    p \equiv -\frac{1}{3} \Delta_{\alpha\beta}T^{\alpha \beta},\quad
    \pi^{\mu\nu} \equiv  \Delta^{\mu\nu}_{\enskip\alpha\beta} T^{\alpha\beta}.
\end{align}
Here $\Delta^{\mu\nu} \equiv g^{\mu\nu}-u^\mu u^\nu$ is a projector which extracts Lorentz vector components transverse to the flow velocity $u^\mu$ and $u^\nu$ and $\Delta^{\mu\nu}_{\enskip\alpha\beta} \equiv \frac{1}{2}(\Delta^\mu_{\enskip\alpha} \Delta^\nu_{\enskip\beta} + \Delta^\nu_{\enskip\alpha} \Delta^\mu_{\enskip\beta}) -\frac{1}{3}\Delta^{\mu\nu}\Delta_{\alpha\beta}$ is also a projector which makes a second rank tensor symmetric, traceless, and transverse to $u^\mu$ and $u^\nu$.

A constitutive relation for $\pi^{\mu \nu}$ up to the second order in derivatives can be written as  \cite{BRSSS}
\begin{align}
    \pi^{\mu\nu} &= 2\eta\sigma^{\mu\nu} - 2\eta  \tau_\pi \left[ \Delta^{\mu\nu}_{\enskip\alpha\beta} D \sigma^{\alpha\beta} + \frac{1}{d-1} \sigma^{\mu\nu} \partial_\alpha u^\alpha \right] \nonumber \\ 
    &- \lambda \Delta^{\mu\nu}_{\enskip\alpha\beta} \sigma^{\alpha}_{\enskip\lambda} \sigma^{\beta \lambda} ,\label{eq:BRSSS}
\end{align}
where $\eta$ is the shear viscosity, $\tau_\pi$ 
 is the relaxation time, $\lambda$ is a transport coefficient at the second order, $\sigma^{\mu\nu} \equiv \Delta^{\mu\nu}_{\enskip\alpha\beta} \partial^\alpha u^\beta$, $d(=4)$ is dimension of the system, and $D \equiv u^\lambda \partial_\lambda$ is the time derivative in the co-moving frame.
We here neglect terms including Riemann and Ricci tensor  and the vorticity which appeared in the original relation \cite{BRSSS}.
In the causal theory of hydrodynamics, dissipative currents  can be promoted to dynamical valuables.
By using the first order relation, $\pi^{\mu\nu} = 2\eta \sigma^{\mu\nu}$, one rewrites Eq.~(\ref{eq:BRSSS}) as
\begin{align}
    \pi^{\mu\nu} = 2\eta\sigma^{\mu\nu} 
    &- \tau_\pi \left[ \Delta^{\mu\nu}_{\enskip\alpha\beta} D \pi^{\alpha\beta} + \frac{d}{d-1} \pi^{\mu\nu} \partial_\alpha u^\alpha \right] \nonumber \\ 
    &- \frac{\lambda}{4\eta^2} \Delta^{\mu\nu}_{\enskip\alpha\beta} \pi^\alpha_{\enskip\lambda}\pi^{\beta\lambda}.\label{eq:BRSSS_resum}
\end{align}
This is the so-called M\"uller--Israel--Stewart type constitutive equation \cite{Muller:1967zza,Israel:1976tn,Israel:1979wp,Hiscock:1983zz} of the second order causal hydrodynamics which we call the Baier--Romatschke--Son--Starinets--Stephanov (BRSSS) equation and employ in the following analysis.

In principle, one might choose any initial conditions for a differential equation of the first order in time.
On the other hand, the shear stress tensor $\pi^{\mu \nu}$ is one of the dissipative currents, in order to be small in comparison with thermodynamic pressure.
Thus it is by no means trivial whether any initial conditions for $\pi^{\mu \nu}$ can be chosen or not.
This is exactly the issue that we are going to address in the next section from a viewpoint of nonlinear causality.

If one considers the conformal system, 
the temperature $T$ is the only dimensional quantity of the system. 
In this specific case, thermodynamic quantities satisfy $e \propto T^4$ and $s \propto T^3$ and the transport coefficients are written as 
\begin{align}
 \eta = C_\eta s, \quad \tau_\pi = \frac{C_{\tau_\pi}}{T}, \quad \lambda = C_{\lambda}\frac{\eta}{T},\label{eq:transport_coeff_confomal}
\end{align}
where $s$ is the entropy density and $C_\eta$, $C_{\tau_\pi}$, and $C_{\lambda}$ are dimensionless parameters which characterize transport properties of the fluid at the second order. 
In the case of $\mathcal{N}=4$ supersymmetric Yang--Mills theory, these coefficients are obtained as \cite{Kovtun:2004de,BRSSS}
\begin{align}
    C_\eta = \frac{1}{4\pi},\quad  C_{\tau_\pi}=\frac{2-\ln 2}{2\pi}, \quad C_{\lambda} = \frac{1}{2\pi}.\label{eq:transport_coeff_AdSCFT}
\end{align}
In this paper, we commonly use these values even when the system is nonconformal, regardless of the model equation of state to be introduced below.

\subsection{Equation of state}
\label{sec:EoS}

In this study, we consider two models of the equation of state (EoS): the equation of state of a conformal system (conformal EoS) and the equation of state from numerical calculations of lattice QCD (lattice EoS).
We employ these two models for the sake of comparison to see effects of nonconformality on final results.

For the conformal system, the trace of the energy-momentum tensor vanishes, $T^\mu_{\enskip \mu} = e-3p = 0$. Therefore the conformal EoS is\footnote{Since this EoS is also obtained by assuming the ideal gas system of relativistic particles, it is sometimes recognized that interactions among constituent particles are neglected in this particular EoS. However, it is not necessary for this assumption to be true. Neglecting interactions among the constituent particles is not necessary to obtain this EoS. Only an assumption that particles are massless leads to the conformal system and, consequently, $p=e/3$.} 
\begin{align}
    p = \frac{1}{3}e.\label{eq:conformal_eos}
\end{align}
In this EoS, the square of sound velocity is $c_s^2 = dp/de = 1/3$.

A realistic EoS of the many body system of quarks and gluons is obtained from the first principle numerical calculations of lattice QCD at finite temperature.
We employ the following parametrization of pressure as a function of temperature  obtained by fitting lattice QCD results \cite{HotQCD:2014kol}:
\begin{align}
    \frac{p}{T^4} = \frac{1}{2}\left\{1+\tanh\left[c_t(\bar{t} - t_0)\right]\right\} \frac{p_{\rm{id}} + a_n/\bar{t} + b_n/\bar{t}^2 + d_n/\bar{t}^4}{1 + a_d/\bar{t} + b_d/\bar{t}^2 + d_d/\bar{t}^4 },\label{eq:lattice_eos}
\end{align}
where $\bar{t} = T/(0.154\enskip\rm{GeV})$ is the dimensionless temperature, $p_{\rm{id}} = 95\pi^2/180$ is a prefactor of pressure for an ideal gas at the Stefan--Boltzmann limit including the degree of freedom for three flavor QCD, and $c_t = 3.8706$, $t_0 = -0.9761$, $a_n = -8.7704$, $b_n = 3.9200$, $d_n = 0.3419$, $a_d = -1.26$, $b_d = 0.8425$, and $d_d = -0.0475$ are the fitting parameters \cite{HotQCD:2014kol}.
Although these parameters are obtained by fitting the lattice QCD results in the range $0.13 < T < 0.4 $ GeV \cite{HotQCD:2014kol}, we utilize them in the whole temperature region.
The other thermodynamic quantities and the square of sound velocity ($c_s^2 = dp/de$) are obtained from Eq.~(\ref{eq:lattice_eos}) through the thermodynamic relations.
In this way, the lattice EoS, $p=p(e)$, is obtained.

\subsection{Equation of Motion in One-Dimensional Expanding System}
\label{sec:EoM}
In what follows, we assume translational and rotational symmetries in the transverse plane perpendicular to the collision axis and boost invariance in the longitudinal direction. Then we employ the boost invariant solution of flow four-velocity \cite{Bjorken:1982qr}, $u^{\mu}_{\mathrm{Bj}} = (t, 0, 0, z)/\tau$, where $\tau = \sqrt{t^2-z^2}$ is the proper time.
From Eqs.~(\ref{eq:energy-momentum-conservation}) and (\ref{eq:BRSSS_resum}), one obtains \cite{BRSSS}
 \begin{align}
    \tau \frac{de}{d\tau} &= -e-p(e) + \phi, \label{eq:Bjorken_eq_for_energy_density}\\
    \tau_\pi \frac{d\phi}{d\tau} &= \frac{4\eta}{3\tau}- \phi-\frac{4}{3}\frac{\tau_\pi}{\tau} \phi  -\frac{\lambda}{2\eta^2} \phi^2, \label{eq:Bjorken_eq_for_constitutive}
 \end{align}
where $\phi \equiv \pi^{00}-\pi^{33}$ is the ``shear pressure".
From the translational and rotational symmetries in the transverse plane, the shear stress tensor for the local rest frame in the one-dimensional boost invariant system is
\begin{align}
\label{eq:eigen_value_pimunu}
    \pi^{\mu\nu}_{\rm{LRF}, \rm{Bj}} = \mathrm{diag} \left(0,\frac{\phi}{2},\frac{\phi}{2},-\phi\right). 
\end{align}
Therefore, longitudinal and transverse pressures including the correction from shear stress tensor become
$p_L=p-\phi$ and $p_T = p+\phi/2$, respectively.
For a given EoS, $p=p(e)$, one solves a set of simultaneous equations (\ref{eq:Bjorken_eq_for_energy_density}) and (\ref{eq:Bjorken_eq_for_constitutive}) numerically. 

These equations (\ref{eq:Bjorken_eq_for_energy_density}) and (\ref{eq:Bjorken_eq_for_constitutive}) are reduced to a single ordinary differential equation in the case of the conformal EoS, which is discussed in details in Appendix \ref{sec:hydrodynamization}.

\subsection{Time Evolution of the Inverse Reynolds Number}
\label{sec:time_evolution_of_Reynolds_no}

\begin{figure}[htpb]
    \centering
    \includegraphics[clip,width=\linewidth]{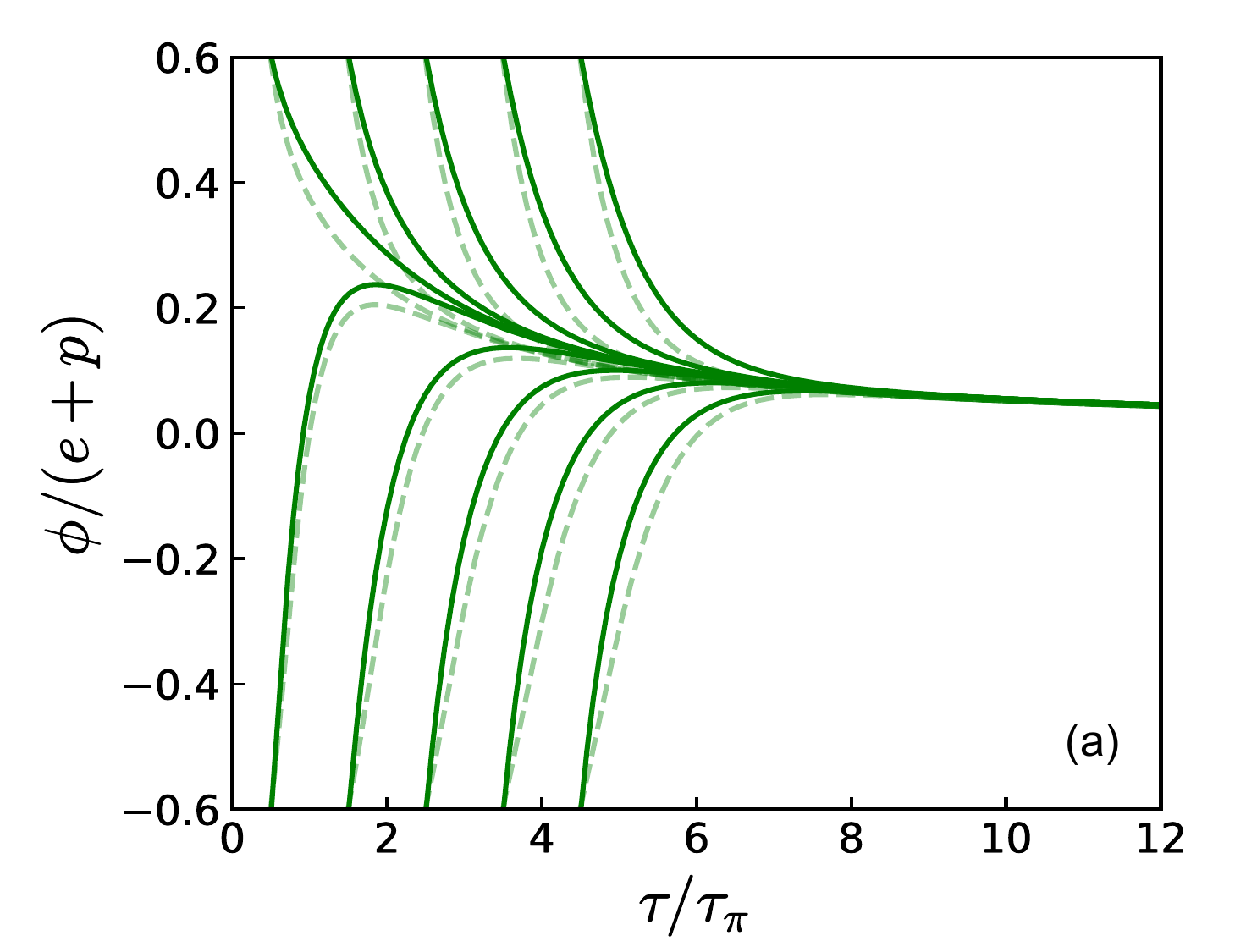}
    \includegraphics[clip,width=\linewidth]{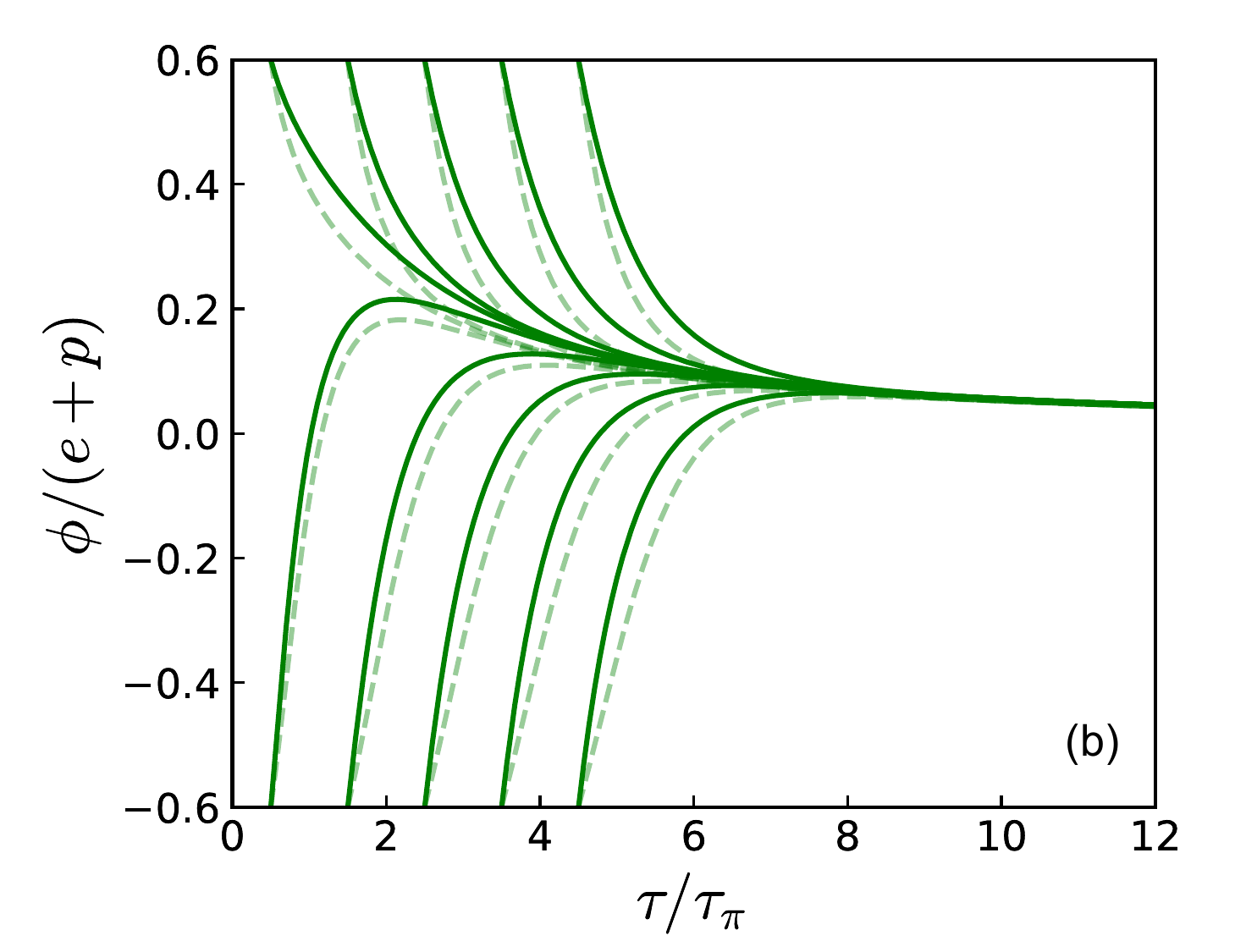}
    \caption{
    Examples of solutions of hydrodynamic equations for several initial conditions when we employ (a) the conformal EoS  and (b) the lattice EoS.
    The solid (dashed) lines represent the solutions with (without) nonlinear $\phi^2$ term in Eq.~(\ref{eq:Bjorken_eq_for_constitutive}).
     }
    \label{fig:time_evolution_of_inverse_Reynolds_no}
\end{figure}
Figure~\ref{fig:time_evolution_of_inverse_Reynolds_no} shows the time evolution of the dimensionless nonequilibrium quantity, $\phi/(e+p)$, from the solution of the hydrodynamic equations (\ref{eq:Bjorken_eq_for_energy_density}) and (\ref{eq:Bjorken_eq_for_constitutive}) for (a) the conformal EoS and (b) the lattice EoS.
Since the nonlinear term ($\propto \phi^2$) in Eq.~(\ref{eq:Bjorken_eq_for_constitutive}) was neglected in several studies (see, e.g., Ref.~\cite{heller2015hydrodynamics}), the solution with this nonlinear term (solid lines) is compared with the one without it (dashed lines). 
Initial times are increased incrementally from $\tau/\tau_{\pi} = 0.5$ to $4.5$ at intervals of 1.0 for $\phi/(e+p) = \pm 0.6$.
In the case of the lattice QCD EoS, one needs to fix the energy scale since it is no longer conformal.
Therefore, as initial conditions, we choose $e = 5$ GeV/fm$^3$, calculate the pressure $p = p(e = 5$ GeV/fm$^3$) and choose the shear pressure so that it becomes $\phi/(e+p) = \pm 0.6$.

The inverse Reynolds number of one-dimensional expanding fluids \cite{Baier:2006um}
can be defined as\footnote{In the literature, one finds various definitions of the inverse Reynolds number which are different from each other by a factor $\mathcal{O}(1)$.}
\begin{align}
    Re^{-1} \equiv \frac{\mid \phi \mid }{e+p}.\label{eq:inverse_Reynolds_no}
\end{align}
Therefore Fig.~\ref{fig:time_evolution_of_inverse_Reynolds_no} shows nothing but the time evolution of the inverse Reynolds number in the one-dimensional expanding fluids \cite{Chattopadhyay:2019jqj}.

It is well-known that the solutions of Eqs.~(\ref{eq:Bjorken_eq_for_energy_density}) and (\ref{eq:Bjorken_eq_for_constitutive}) rapidly converge to a single curve which is sometimes called ``hydrodynamic attractor" \cite{heller2015hydrodynamics,Romatschke:2016hle,Romatschke:2017vte,Florkowski:2017olj,Jankowski:2023fdz}.
It is obvious from Fig.~\ref{fig:time_evolution_of_inverse_Reynolds_no} that all results exhibit this behavior, regardless of whether one includes the nonlinear term in Eq.~(\ref{eq:Bjorken_eq_for_constitutive}).
In fact, in Appendx\ref{sec:hydrodynamization} we also exhibit some solutions that move away from the hydrodynamic attractor if we start from too small $\phi/(e+p)$.

\section{Results} \label{sec:results}
We first obtain nonlinear causal conditions in one-dimensional expanding system within the BRSSS equation in Sec.~\ref{sec:nonlinear_causality_in_1D}.
For given transport coefficients from AdS/CFT, the necessary and sufficient conditions can be translated into the constraints on the dimensionless nonequilibrium quantity (or the inverse Reynolds number) in Sec.~\ref{sec:constraint_Re-1}.
Even if the system starts from the local equilibrium state, the system can deviated from it due to strong longitudinal expansion and may have a large value of the inverse Reynolds number.
This means the initial conditions can be chosen so that the system obeys the causality even during its evolution. This will be done in Sec.~\ref{sec:constraint_IC}.
In Sec.~\ref{sec:Results_from_lattice_EoS}, we discuss how the results with the conformal EoS are modified when we employ the lattice EoS.

\subsection{Nonlinear Causality in One-Dimensional Expanding System}
\label{sec:nonlinear_causality_in_1D}

The necessary and the sufficient conditions for the relativistic fluids to be causal have been obtained in the nonlinear regime of relativistic hydrodynamic equations \cite{bemfica2021nonlinear}.
These conditions are written in terms of thermodynamic variables, 
dissipative quantities, 
and transport coefficients. 
We translate these necessary and sufficient conditions into the ones specifically in the one-dimensional expanding system which obeys the BRSSS equation (\ref{eq:BRSSS_resum}).

The eigenvalues of the shear stress tensor, $\pi^\mu_{\enskip \nu}$, are defined as $\Lambda_\alpha$ ($\alpha = 0$, $1$, $2$, $3$). 
These eigenvalues obey $\Lambda_0 = 0$ from $\pi^{\mu}_{\enskip \nu} u^\nu = 0$ and $\Lambda_1 + \Lambda_2 + \Lambda_3 = 0$ from  $\pi^{\mu}_{\enskip \mu} =0 $.
Without loss of generality, we set $\Lambda_1 \le \Lambda_2 \le \Lambda_3$ with $\Lambda_1 \le 0 \le \Lambda_3$.
In the local rest frame of one-dimensional expanding system, the shear stress tensor is already diagonalized as Eq.~(\ref{eq:eigen_value_pimunu}).
When the shear pressure $\phi$ is positive (negative), one can assign that $\Lambda_0 = 0$, $\Lambda_1 = \Lambda_2 = -\phi/2$, and $\Lambda_3 = \phi$ \  ($\Lambda_0 = 0$, $\Lambda_1 = \phi$, and $\Lambda_2 = \Lambda_3 = -\phi/2$).

The necessary conditions of one-dimensional expanding fluids from the BRSSS equation (\ref{eq:BRSSS_resum}) are summarized as follows:
\begin{align}
&    \eta \ge 0, \quad \frac{\eta}{\tau_{\pi}} \ge 0, \quad e+p-\frac{\eta}{\tau_\pi} \ge 0, \label{eq:Necessary_BRSSS1}\\
& e+p -\frac{\phi}{2} -\frac{\eta}{\tau_{\pi}} \ge 0,\\
& e+p + \phi -\frac{\eta}{\tau_{\pi}} \ge 0, \label{eq:Necessary_BRSSS2}\\
&  \left(e+p-\frac{\phi}{2}\right)c_s^2 + \frac{4}{3}\left(-\frac{\phi}{2}\right)+  \frac{4\eta}{3\tau_\pi}   \ge 0, \label{eq:Necessary_BRSSS3}\\
&    (e+p+\phi)c_s^2 + \frac{4}{3}\phi+ \frac{4\eta}{3\tau_\pi}    \ge 0, \label{eq:Necessary_BRSSS4}\\
&    \left(e+p-\frac{\phi}{2}\right)(1- c_s^2) - \frac{4\eta}{3\tau_\pi} + \frac{2}{3}\phi \ge 0, \label{eq:Necessary_BRSSS5}\\
&    (e+p+\phi)(1- c_s^2) - \frac{4\eta}{3\tau_\pi} - \frac{4}{3}\phi \ge 0. \label{eq:Necessary_BRSSS6}
\end{align}

All of these inequalities must be satisfied so that the one-dimensional expanding fluids obey causality. 
When even one of them is not satisfied, it means the system violates causality.
It should be also kept in mind that, even if all the necessary conditions are satisfied, it does not mean that causality is guaranteed \cite{bemfica2021nonlinear}.

The sufficient conditions of one-dimensional expanding fluids from the BRSSS equation (\ref{eq:BRSSS_resum}) depend on the sign of the shear pressure $\phi$.
Regardless of the sign of $\phi$, almost trivial inequalities of the EoS and the transport coefficients are obtained from the sufficient conditions:
\begin{align}
    \eta \ge 0,\quad  \tau_{\pi} \ge 0, \quad  c_s^2 \ge 0.
\end{align}
On the other hand, the nontrivial sufficient conditions depending on the sign of $\phi$ are summarized as follow:

In the case of $\phi>0$, 
\begin{align}
\label{eq:suffficient_positive_phi_first}
    &e+p- \frac{\phi}{2} - \frac{\eta}{\tau_\pi} \ge 0,\\
    &\frac{4}{3}\left(\phi+\frac{\eta}{\tau_\pi} \right) + \left(\frac{1}{2} +  c_s^2\right) \phi+ \frac{3 c_s^2 \phi^2 }{e+p-\frac{\phi}{2} - \frac{\eta}{\tau_\pi}} \nonumber \label{upperBorder_suf_conformal}\\
    &  \le (e+p)(1-c_s^2), \\
    &\left(e+p-\frac{\phi}{2}\right)c_s^2 - \frac{2}{3}\phi+\frac{\eta}{3\tau_\pi}  \ge 0, \\
    &\left(\frac{\eta}{\tau_\pi}\right)^2- 3c_s^2 \phi^2\ge 0, \\
    &\left(e+p-\frac{\phi}{2}\right)c_s^2+\frac{4}{3}\left(-\frac{\phi}{2}+\frac{\eta}{\tau_\pi}\right)   \nonumber\\
\label{eq:suffficient_positive_phi_last}
    &  \ge \frac{(e+p+\phi)\left(e+p-\frac{\phi}{2}+\frac{2\eta}{\tau_\pi}\right)}{3\left(e+p-\frac{\phi}{2}\right)}.
\end{align}

In the case of $\phi<0$, 
\begin{align}
\label{eq:suffficient_negative_phi_first}
    &e+p+\phi - \frac{\eta}{\tau_\pi} \ge 0, \\
    &\frac{4}{3}\left(-\frac{\phi}{2}+\frac{\eta}{\tau_\pi} \right) - \left(1+\frac{1}{2}c_s^2\right)\phi  + \frac{3c_s^2\phi^2}{e+p+\phi - \frac{\eta}{\tau_\pi}} \nonumber\\
    &  \le (e+p)(1-c_s^2), \\
    &(e+p+\phi)c_s^2 + \frac{4}{3}\phi +\frac{\eta}{3\tau_\pi} \ge 0, \\
    &\left(\frac{\eta}{\tau_\pi}\right)^2 - 3 c_s^2 \phi^2 \ge 0, \\
\label{eq:suffficient_negative_phi_last}
    &\frac{4}{3}(\phi + \frac{\eta}{\tau_\pi}) + (e+p+\phi)c_s^2 \nonumber \\
    &\ge \frac{\left(e+p-\frac{\phi}{2}\right)^2\left(e+p+\phi+\frac{2\eta}{\tau_\pi}\right)}{3(e+p+\phi)^2} .
\end{align}

When all of these inequalities (Eqs.~(\ref{eq:suffficient_positive_phi_first})-(\ref{eq:suffficient_positive_phi_last}) for positive $\phi$ or Eqs.~(\ref{eq:suffficient_negative_phi_first})-(\ref{eq:suffficient_negative_phi_last}) for negative $\phi$) are satisfied at the same time, the evolution of system is guaranteed to be causal.
See also Appendix \ref{sec:causality appendix} for more details on the necessary and the sufficient conditions shown above.

\subsection{Constraint on the inverse Reynolds number}
\label{sec:constraint_Re-1}

By employing conformal EoS and transport coefficients from AdS/CFT, the causality conditions are  written as inequalities among thermodynamic variables and the shear pressure:\footnote{Although the pressure $p$ is just equal to $e/3$ in the conformal EoS, we still keep $p$ in these inequalities since we compare these results with the one from lattice EoS in Sec.~\ref{sec:Results_from_lattice_EoS}.}
\footnote{Our result is consistent with the one obtained in Ref.~\cite{bemfica2021nonlinear}: $-4/5 <\Lambda_{a}/(e+p) \le 3/5$ with a different choice of relaxation time, $\tau_\pi T = 5 \eta/s$. }
\begin{align}
    & -0.47 \le \frac{\phi}{e+p} \le 0.23, \quad (\mbox{necessary}) \label{eq:Necessary_Bj}\\
    & -0.07 \le \frac{\phi}{e+p} \le 0.07. \quad (\mbox{sufficient})\label{eq:Sufficient_Bj}
\end{align}
For the necessary conditions, Eqs.~(\ref{eq:Necessary_BRSSS5}) and (\ref{eq:Necessary_BRSSS6}) provide the constraint (\ref{eq:Necessary_Bj}) for $\phi<0$ and $\phi>0$, respectively.
On the other hand, the constraint (\ref{eq:Sufficient_Bj}) comes from Eqs.~(\ref{upperBorder_suf_conformal}) and (\ref{eq:suffficient_negative_phi_last}).
See also Appendix \ref{sec:causality appendix} to understand how each necessary condition is obtained from a discussion of characteristic velocity. 

\begin{figure}[htpb]
    \centering
    \includegraphics[clip,width=\linewidth]{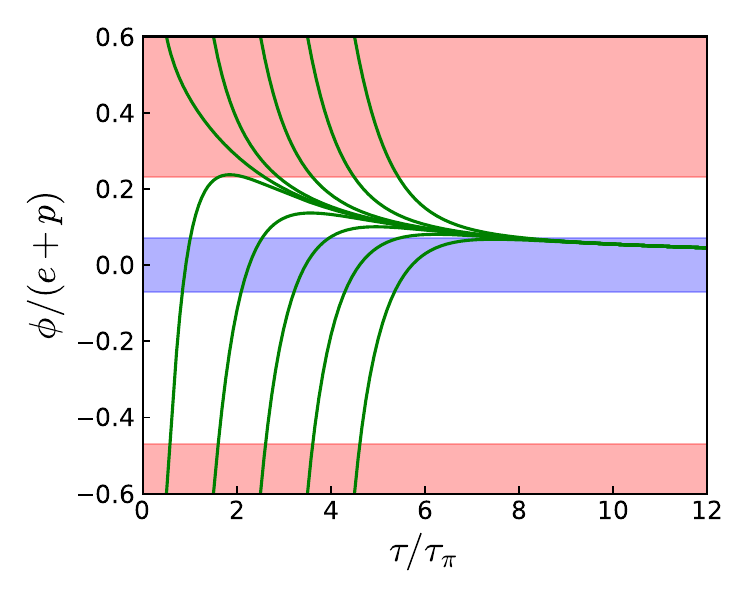}
    \caption{
    Examples of the solutions of hydrodynamic equations (\ref{eq:Bjorken_eq_for_energy_density}) and (\ref{eq:Bjorken_eq_for_constitutive})  for several initial conditions and acausal (red) and causal (blue) regions from the necessary (\ref{eq:Necessary_Bj}) and the sufficient (\ref{eq:Sufficient_Bj}) conditions, respectively. Note that, in between one cannot determine whether a solution is causal or not in between only from these conditions.
     }
    \label{fig:result1}
\end{figure}

To see how the solutions for hydrodynamic equations of one-dimensional expanding fluids obey or violate the causality, the regions from these conditions (\ref{eq:Necessary_Bj}) and (\ref{eq:Sufficient_Bj}) together with ten examples of solutions are shown in Fig.~\ref{fig:result1}.
When a solution of Eqs.~(\ref{eq:Bjorken_eq_for_energy_density}) and (\ref{eq:Bjorken_eq_for_constitutive}) goes through the acausal region, it does not obey at least one necessary condition and, consequently, violates the relativistic causality.
On the other hand,  when a solution keeps its trajectory in the causal region during the time evolution, it obeys the all sufficient conditions (needless to say, the all necessary conditions, too)  and, consequently, the causality of the dynamics is guaranteed.
With only these necessary and sufficient conditions of nonlinear causality, it is not yet determined in between whether it is causal or not: All necessary conditions are satisfied, but some of the sufficient conditions are not satisfied.
In any case, the dynamics of fluids tends to be acausal when the inverse Reynolds number is large and the system is far away from local equilibrium.

\subsection{Constraint on Initial Conditions}
\label{sec:constraint_IC}

\begin{figure}[htpb]
    \centering
    \includegraphics[clip,width=\linewidth]{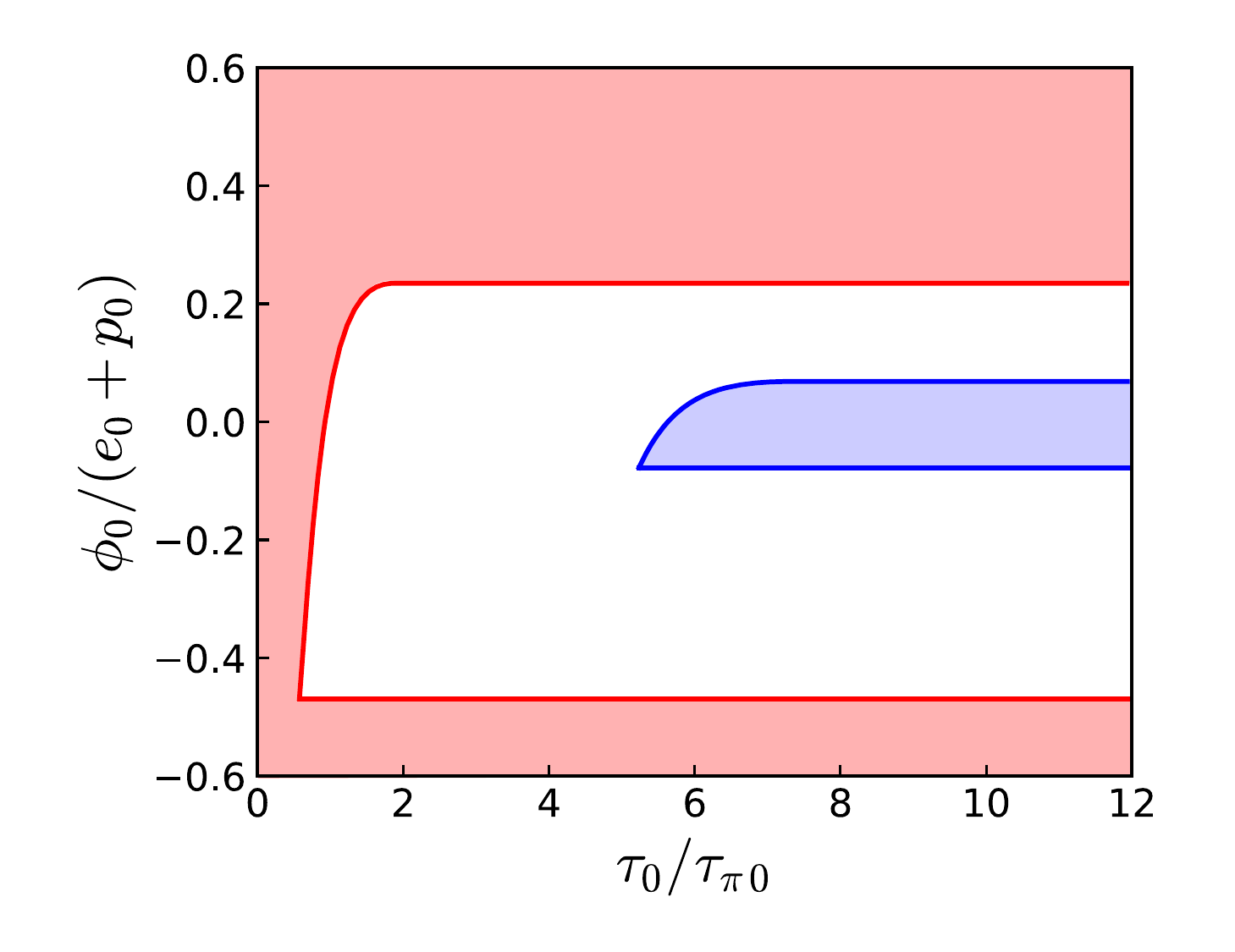}
    \caption{
    The regions of acausal (red) and causal (blue) initial conditions in the $\tau_0/\tau_{\pi0}$-$\phi_0/(e_0+p_0)$ plane. See the text for details.
     }
    \label{fig:result2}
\end{figure}

Any solutions that pass through the acausal regions are by no means acceptable from a viewpoint of relativistic theory.
The initial values of the shear pressure $\phi_0$, the energy density $e_0$, and the resulting pressure $p_0$ calculated from the EoS must be chosen so that the solutions never intrude the acausal regions. 
Thus the initial conditions are considerably constrained in the $\tau_0/\tau_{\pi0}$-$\phi_0/(e_0+p_0)$ plane from the causality argument as shown in Fig.~\ref{fig:result2}.
Wherever an initial value is taken from the region of acausal initial conditions in Fig.~\ref{fig:result2}, the solution of Eqs.~(\ref{eq:Bjorken_eq_for_energy_density}) and (\ref{eq:Bjorken_eq_for_constitutive}) definitely violates causality at least during its evolution.
On the other hand, wherever an initial value is taken from the region of causal initial conditions in Fig.~\ref{fig:result2}, the solution is always guaranteed to be causal during its evolution.

\begin{figure}[htpb]
    \centering
    \includegraphics[clip,width=\linewidth]{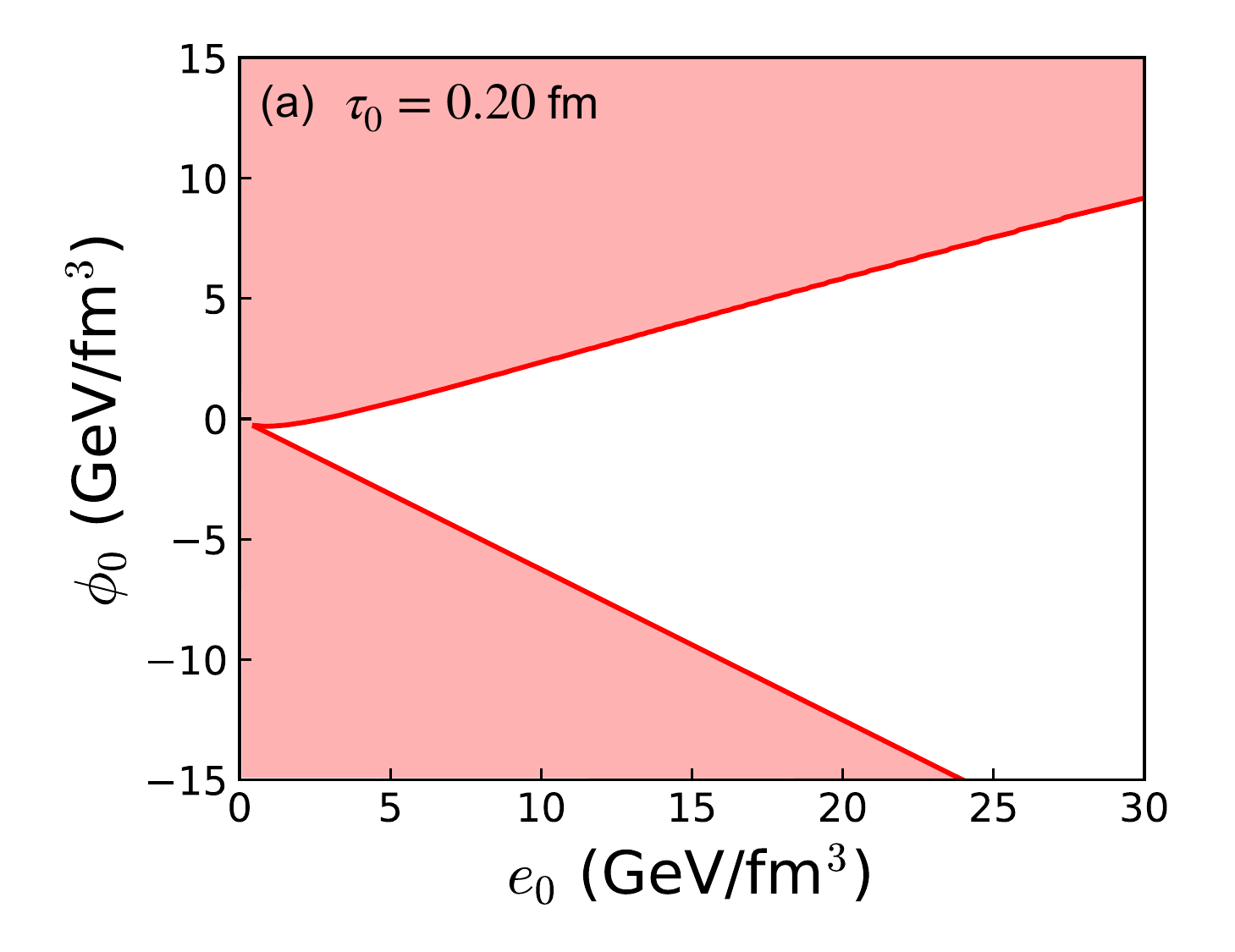}
    \includegraphics[clip,width=\linewidth]{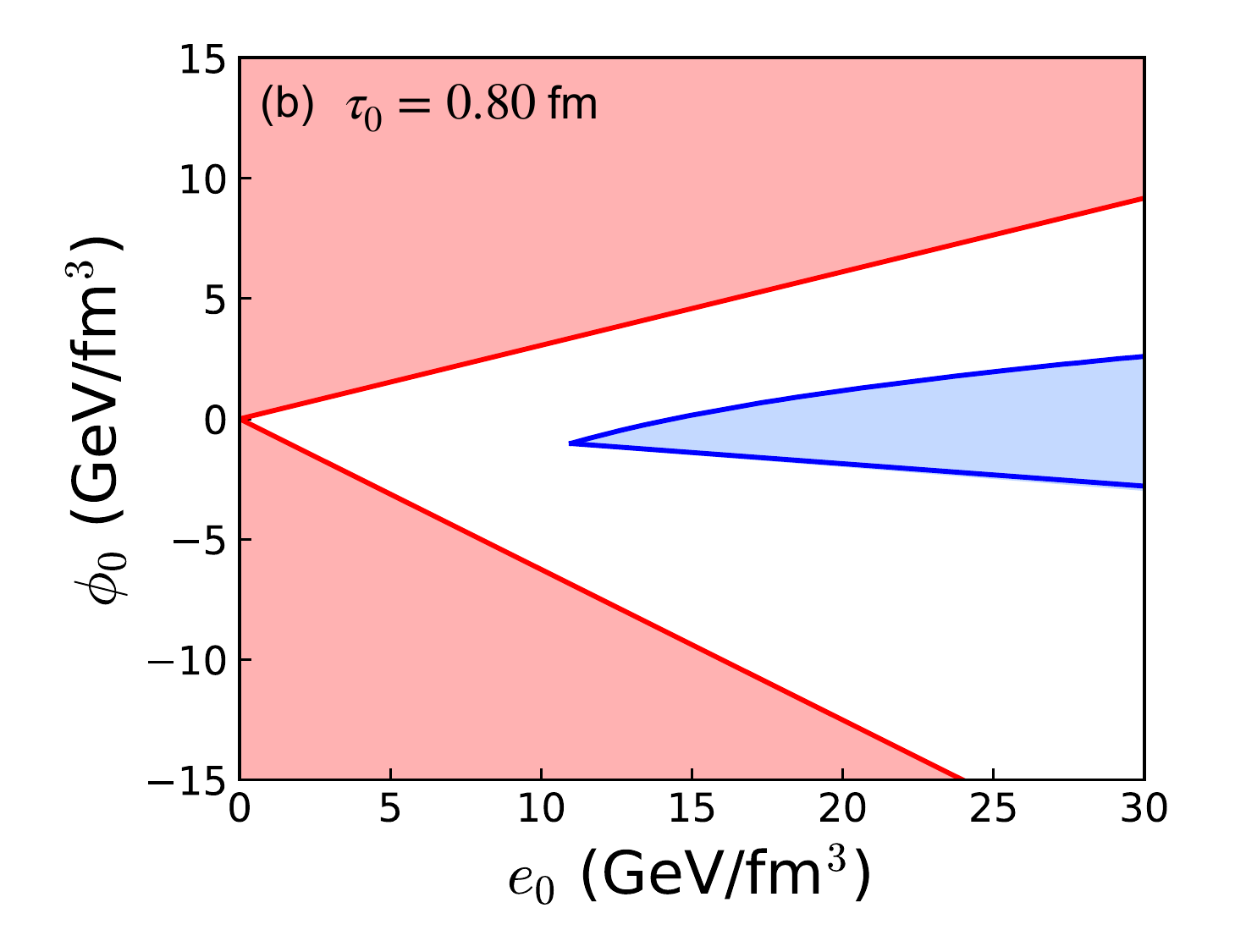}
    \includegraphics[clip,width=\linewidth]{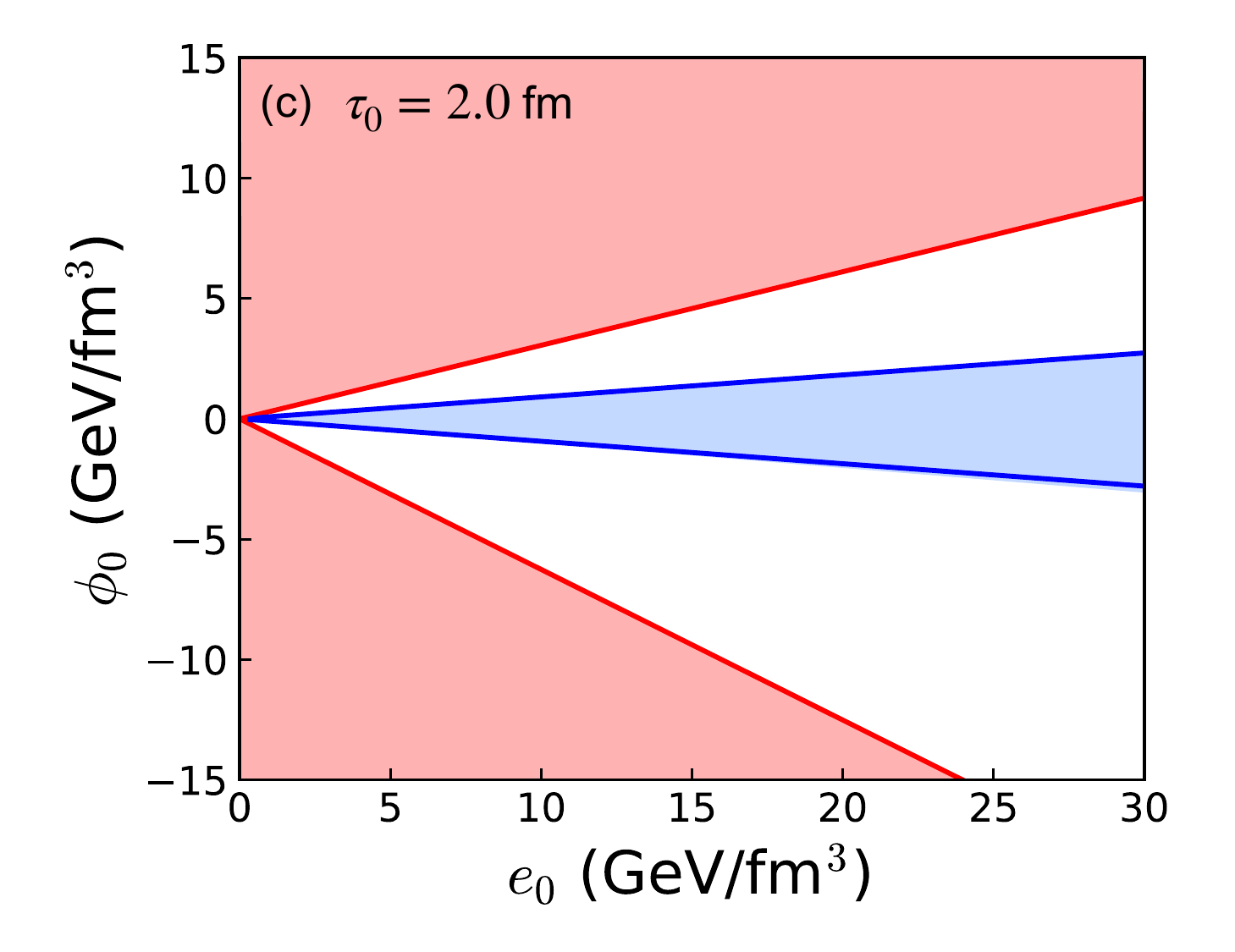}
    \caption{
    The regions of acausal (red) and causal (blue) initial conditions in the $e_0$-$\phi_0$ plane at the initial time, (a) $\tau_0  = 0.2$, (b) $0.8$, and (c) $2.0$ fm. See the text for details.
    }
    \label{fig:result3}
\end{figure}

It would be instructive to draw the regions of initial conditions whose solutions are causal or acausal in another plane.
For a fixed initial proper time $\tau_0 = 0.2$, $0.8$, or $2.0$ fm, the regions of acausal and causal initial conditions are shown in Fig.~\ref{fig:result3}.
For an extremely early initial proper time ($\tau_0 = 0.2$ fm), there is no room for solutions that are guaranteed to be causal, as shown in Fig.~\ref{fig:result3} (a).
Even if the fluids start from a local equilibrium state, the expansion scalar in a one-dimensional expanding system, $\partial_\mu u_{\mathrm{Bj}}^\mu = 1/\tau$, becomes tremendously large for extremely early initial proper time and, consequently, fluids are quickly far away from the local equilibrium state.
As $\tau_{0}$ increases to 0.8 [Fig.~\ref{fig:result3} (b)] and 2.0 fm [Fig.~\ref{fig:result3} (c)], 
the region of causal initial conditions expands. 

\begin{figure}[htpb]
    \centering
    \includegraphics[clip,width=\linewidth]{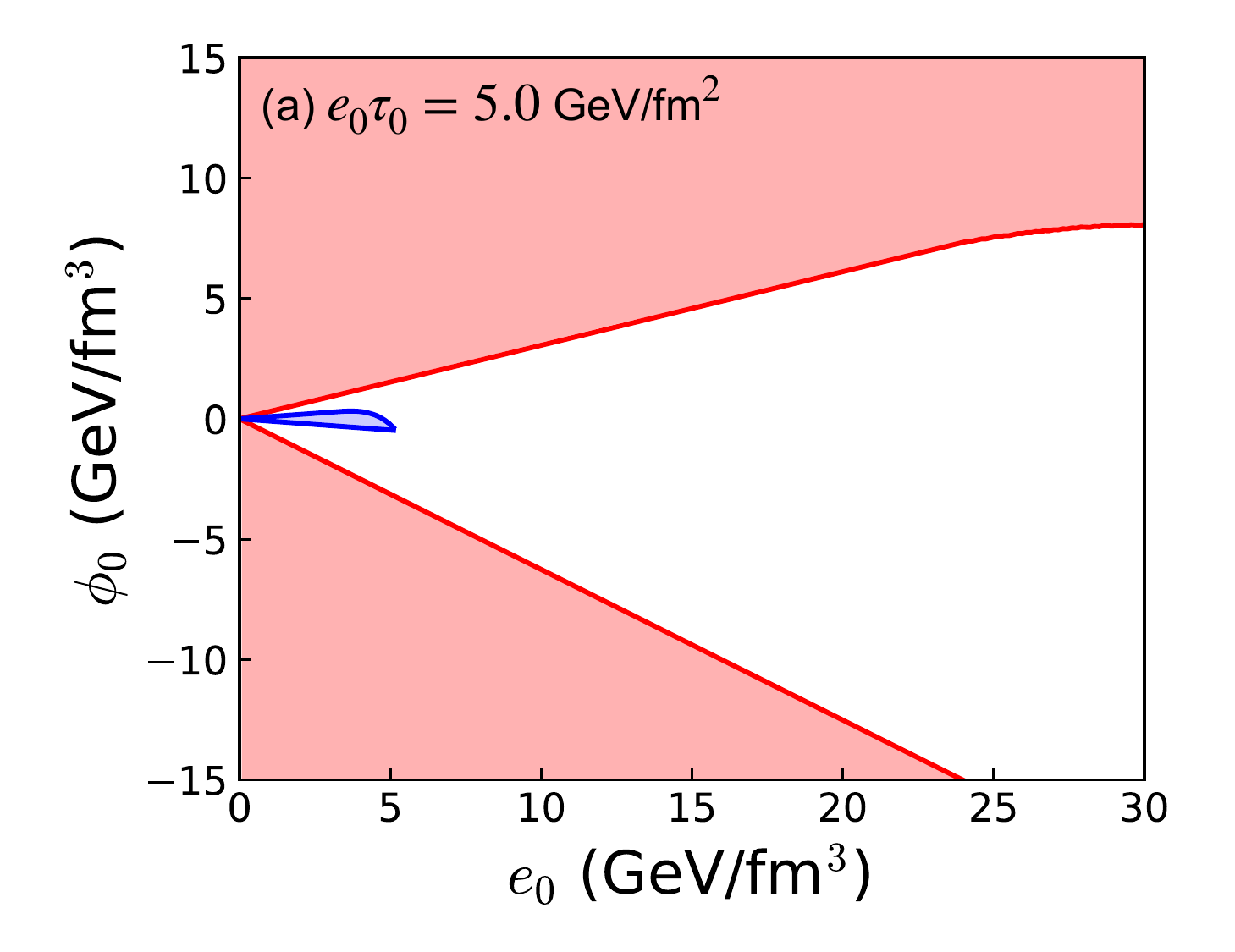}
    \includegraphics[clip,width=\linewidth]{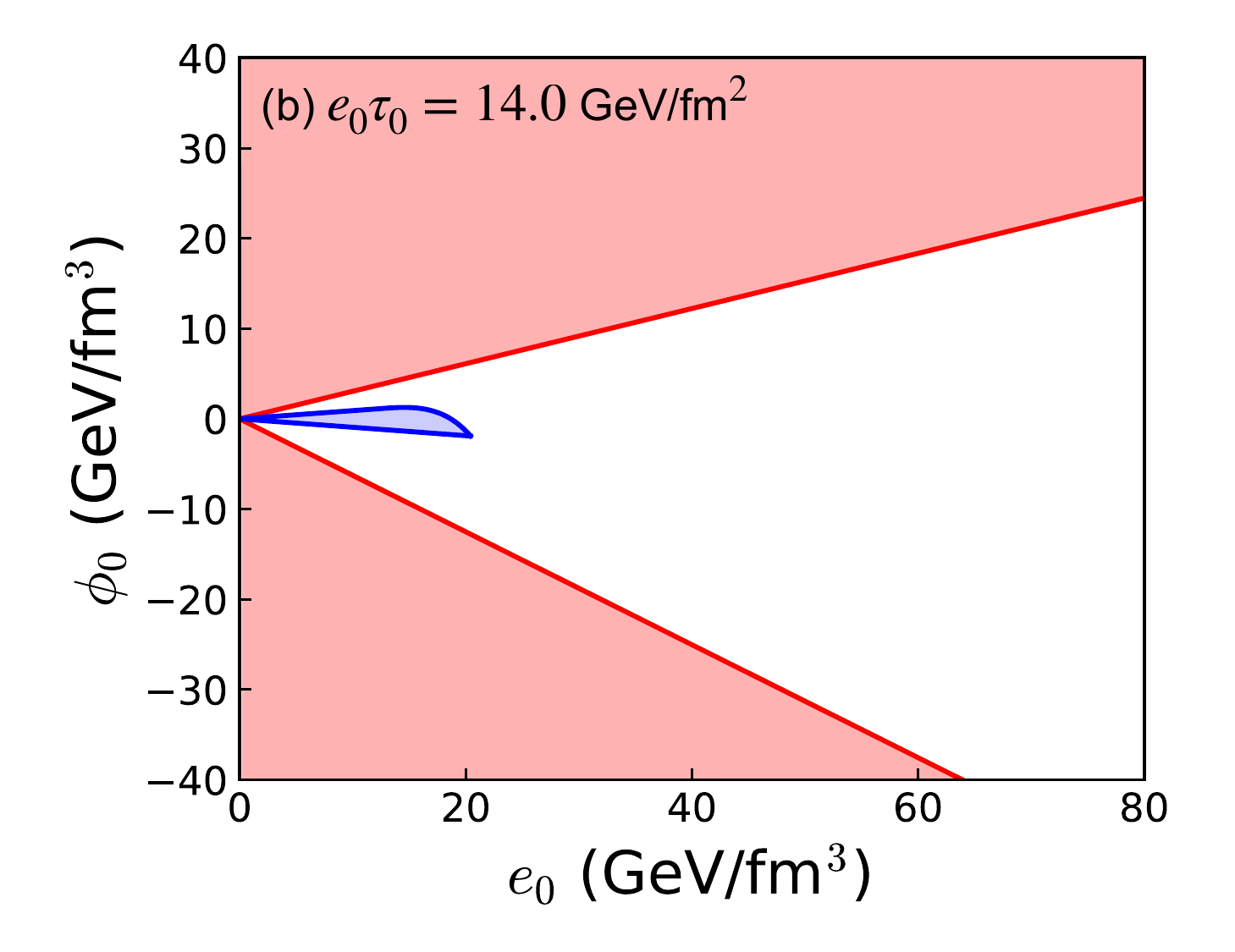}
    \caption{
    The regions of acausal (red) and causal (blue) initial conditions in central heavy-ion collisions at RHIC and LHC energies.
    From the measured transverse energy in central collisions at midrapidity, we fix  (a) $e_0 \tau_0 = 5$ GeV/fm$^2$ at RHIC \cite{STAR:2004moz} and (b) $14$ GeV/fm$^2$ at LHC energies \cite{CMS:2012krf}. 
     }
    \label{fig:result4}
\end{figure}

By combining the results from nonlinear causality with experimental data,  the initial conditions can be further constrained.
From the transverse energy per rapidity, $dE_{\mathrm{T}}/dy$, measured in relativistic heavy-ion collisions,
the Bjorken energy density  \cite{Bjorken:1982qr} can be estimated as
\begin{align}
    e_{0} = \frac{1}{\tau_0 S}\frac{dE_{\mathrm{T}}}{dy},
    \label{eq:Bjorken_energy_density}
\end{align}
where $S$ is an effective area of the overlap region in the transverse plane. 
From Eq.~(\ref{eq:Bjorken_energy_density}), one obtains $e_0 \tau_0 \approx 5$ GeV/fm$^2$ in top 5\% Au+Au collisions at $\sqrt{s_{NN}} = 200$ GeV  \cite{STAR:2004moz} and $\approx 14$ GeV/fm$^2$ in top 5\% Pb+Pb collisions at $\sqrt{s_{NN}}=2.76$ TeV \cite{CMS:2012krf}.
The regions of acausal and causal initial conditions with fixed $e_{0}\tau_0$ are shown in Fig.~\ref{fig:result4}.
One finds from Fig.~\ref{fig:result4} that the acceptable region for a system guaranteed to be causal is highly constrained from experimental inputs.
A point at the maximum energy density  in the region guaranteed to be causal in Fig.~\ref{fig:result4} corresponds to the minimum initial proper time, $\tau_0/\tau_{\pi}$, in Fig.~\ref{fig:result2}. 
Therefore the minimum initial time is obtained from a combination of nonlinear causality and the measured transverse energy.

From the measured transverse energy and the effective transverse area, one obtains
\begin{align}
    e \tau = \frac{1}{S}\frac{dE_{\mathrm{T}}}{dy} & \equiv E_0 \enskip (\mathrm{GeV/fm}^2) = \frac{E_0}{\hbar c} \enskip (\mathrm{1/fm}^3). \label{eq:e0tau0}
\end{align}
By assuming a three-flavor massless ideal gas for the QGP ($N_f = 3$ and $d_{\mathrm{eff}} = 47.5$), the energy density can be transformed into the temperature:
\begin{align}
    e = d_{\mathrm{eff}}\frac{\pi^2}{30}T^4.\label{eq:ideal_gas_qgp}
\end{align}
One can read out the minimum initial time normalized by the relaxation time from Fig.~\ref{fig:result2} as
\begin{align}
    \frac{\tau_{0,\mathrm{min}}}{\tau_\pi} = \frac{\tau_{0,\mathrm{min}} T_{0,\mathrm{max}}}{C_{\tau_{\pi}}} \approx 5.3. \label{eq:normalized_initial_time}
\end{align}
From Eqs.~(\ref{eq:e0tau0})-(\ref{eq:normalized_initial_time}), one finally obtains
\begin{align}
    \tau_{0,\mathrm{min}} \approx 1.6 \times E_0^{-\frac{1}{3}} \enskip (\mathrm{fm}). \label{eq:Hoshino-Hirano_formula} 
\end{align}
This is the minimum initial proper time of conformal fluids with $N_f = 3$ obtained from nonlinear causality and the measured transverse energy. 
It is noted that the power $-1/3$ in Eq.~(\ref{eq:Hoshino-Hirano_formula}) comes from dimensionality in the conformal system and that the prefactor is a nontrivial consequence from nonlinear causality.

From Eqs.~(\ref{eq:Bjorken_energy_density}) and (\ref{eq:Hoshino-Hirano_formula}), maximum energy density and minimum initial proper time are $e_{0,\mathrm{max}} \approx 5$ GeV/fm$^3$ and $\tau_{0,\mathrm{min}} \approx 1$ fm in top 5\% Au+Au collisions at $\sqrt{s_{NN}} = 200$ GeV and $e_{0,\mathrm{max}} \approx 20$ GeV/fm$^3$ and $\tau_{0,\mathrm{min}} \approx 0.7$ fm in top 5\% Pb+Pb collisions at $\sqrt{s_{NN}}=2.76$ TeV.
If one relies strictly on  the sufficient conditions from nonlinear causality in one-dimensional expansion, these impose the minimum bound of the initial proper time.
However, if one imposes the necessary conditions only, the broader regions are available for the initial conditions in one-dimensional expansion of the conformal fluids and the earlier initial time could be acceptable.

\subsection{Results with lattice EoS}
\label{sec:Results_from_lattice_EoS}

For a more realistic analysis, results with the conformal EoS are compared in this subsection with the ones using the lattice EoS.
Note that we replace the model EoS only and keep the equation of motion and the transport coefficients as in the previous subsections.

Since the system is no longer a conformal one, the solutions of Eqs.~(\ref{eq:Bjorken_eq_for_energy_density}) and (\ref{eq:Bjorken_eq_for_constitutive}) depend on the energy scale such as initial temperature or initial energy density.
Thus, we choose $e_0=5$ and $14$ GeV/fm$^3$ at RHIC and LHC energies, respectively, as the typical energy densities.
From the causality conditions for these two energy scales, we obtain the inequalities among thermodynamic variables and the shear pressure as follows: 
In the case of $e_0=5$ GeV/fm$^3$,
\begin{align}
   & -0.48 \le \frac{\phi}{e+p} \le 0.39, \quad (\mbox{necessary})\\
   & -0.05 \le \frac{\phi}{e+p} \le 0.09, \quad(\mbox{sufficient})
\end{align}
while in the case of $e_0 = 14$ GeV/fm$^3$,
\begin{align}
   & -0.49 \le \frac{\phi}{e+p} \le 0.33, \quad (\mbox{necessary})\\
   & -0.06 \le \frac{\phi}{e+p} \le 0.08. \quad (\mbox{sufficient})
\end{align}

\begin{figure}[htpb]
    \centering
    \includegraphics[clip,width=\linewidth]{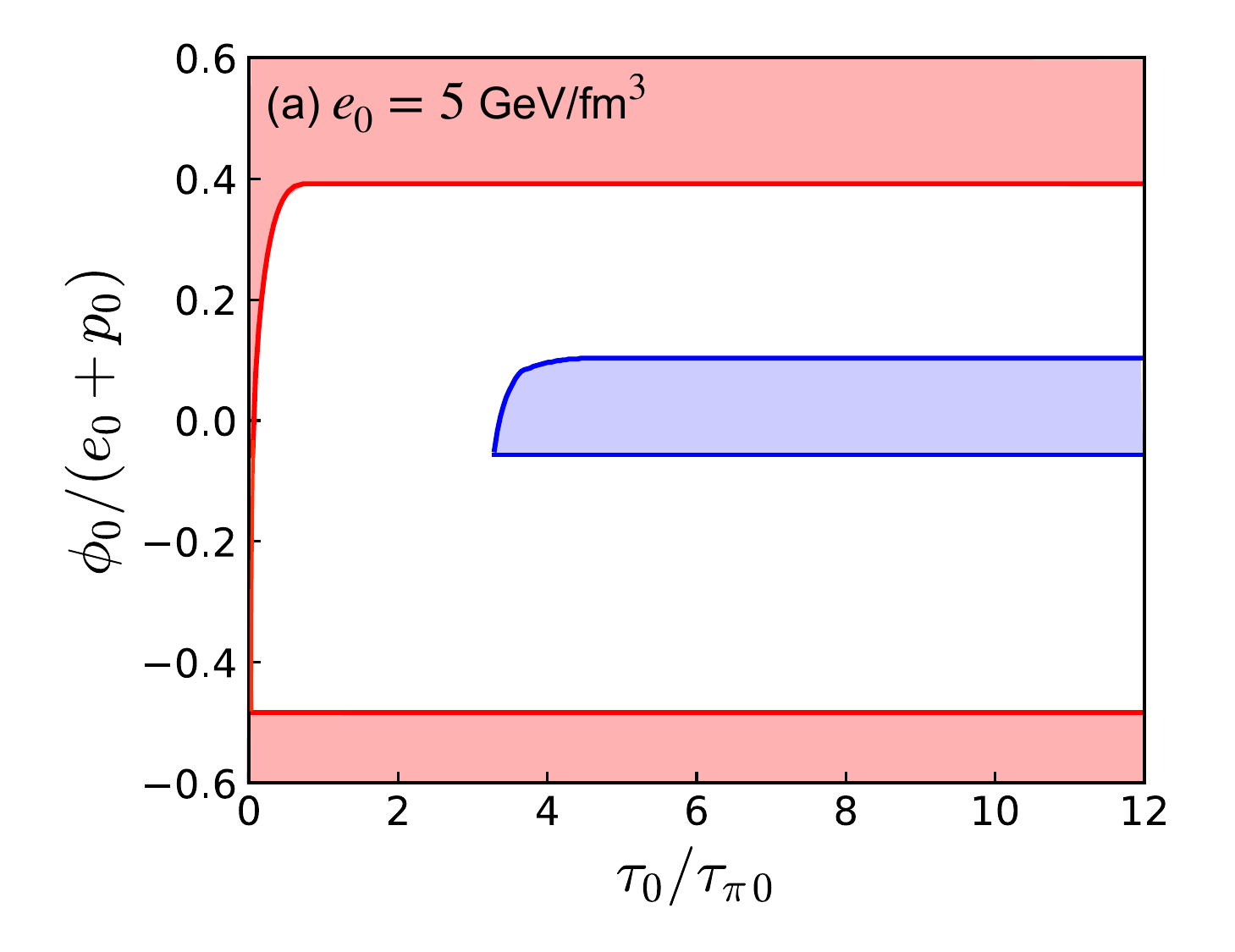}
    \includegraphics[clip,width=\linewidth]{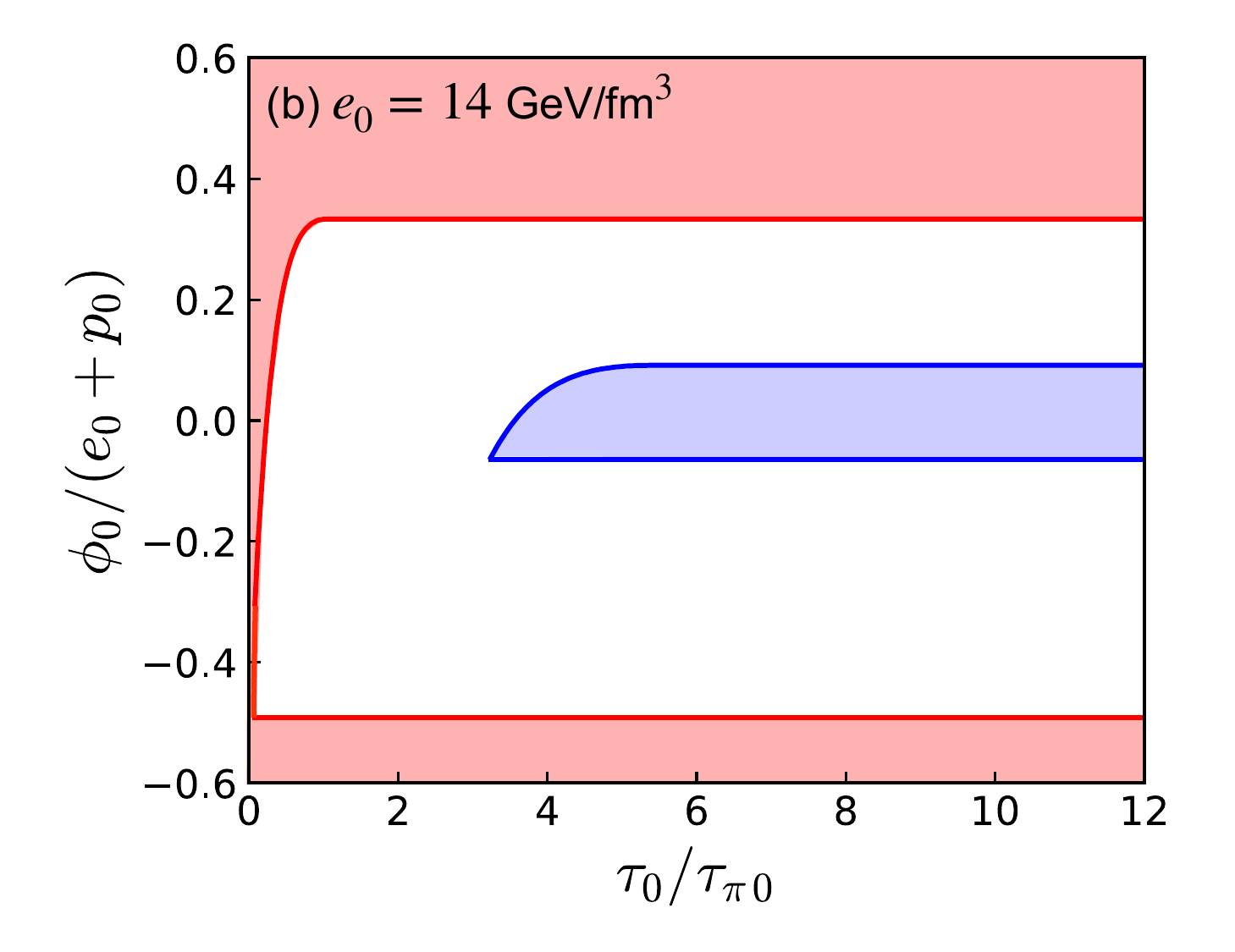}
    \caption{
    The same as Fig.~\ref{fig:result2} but employing the lattice EoS.
    We set (a) $e_0 = 5$ GeV/fm$^3$ and (b) $e_0 = 14$ GeV/fm$^3$.
     }
    \label{fig:result5}
\end{figure}

Figure \ref{fig:result5} shows the regions of acausal and causal initial conditions at (a) $e_0 = 5$ GeV/fm$^3$ and (b) $14$ GeV/fm$^3$, respectively.
By comparison of Fig.~\ref{fig:result5} with Fig.~\ref{fig:result2}, the necessary conditions seem to be relaxed.
One of the  main reasons would be that the square of sound velocity in the lattice EoS is smaller than that in the conformal EoS ($c_s^2=1/3$).
Among the necessary conditions, Eqs.~(\ref{eq:Necessary_BRSSS5}) and (\ref{eq:Necessary_BRSSS6}) set the boundary of whether a solution for a set of initial conditions ($\tau_0$, $\phi_{0}$) violates the causality.
From these equations, it can be said that the smaller the square of sound velocity is, the more relaxed the necessary conditions are. 
Regarding the sufficient conditions, it is not trivial how the sufficient conditions are affected by changing the sound velocity.

\begin{figure}[htpb]
    \centering
    \includegraphics[clip,width=\linewidth]{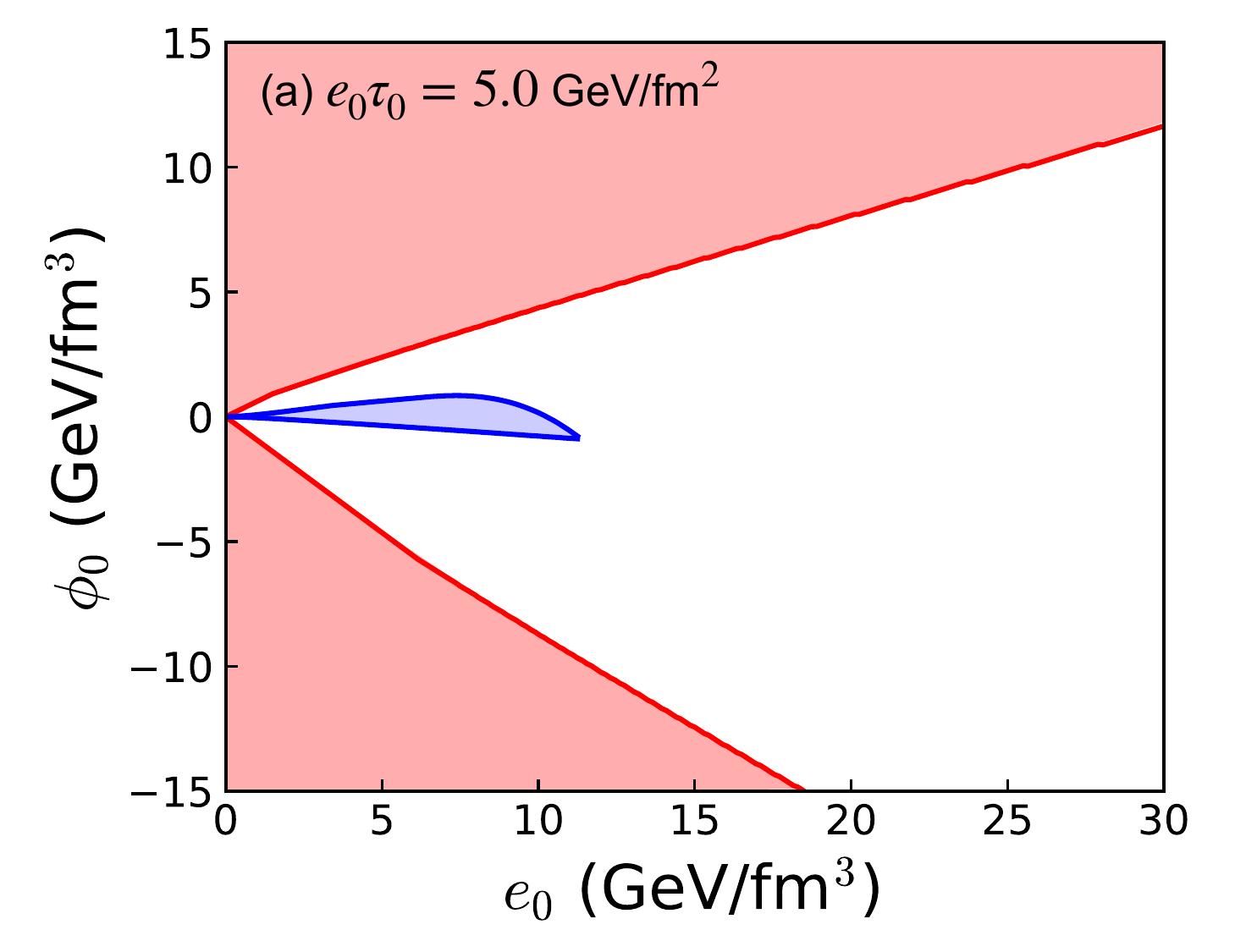}
    \includegraphics[clip,width=\linewidth]{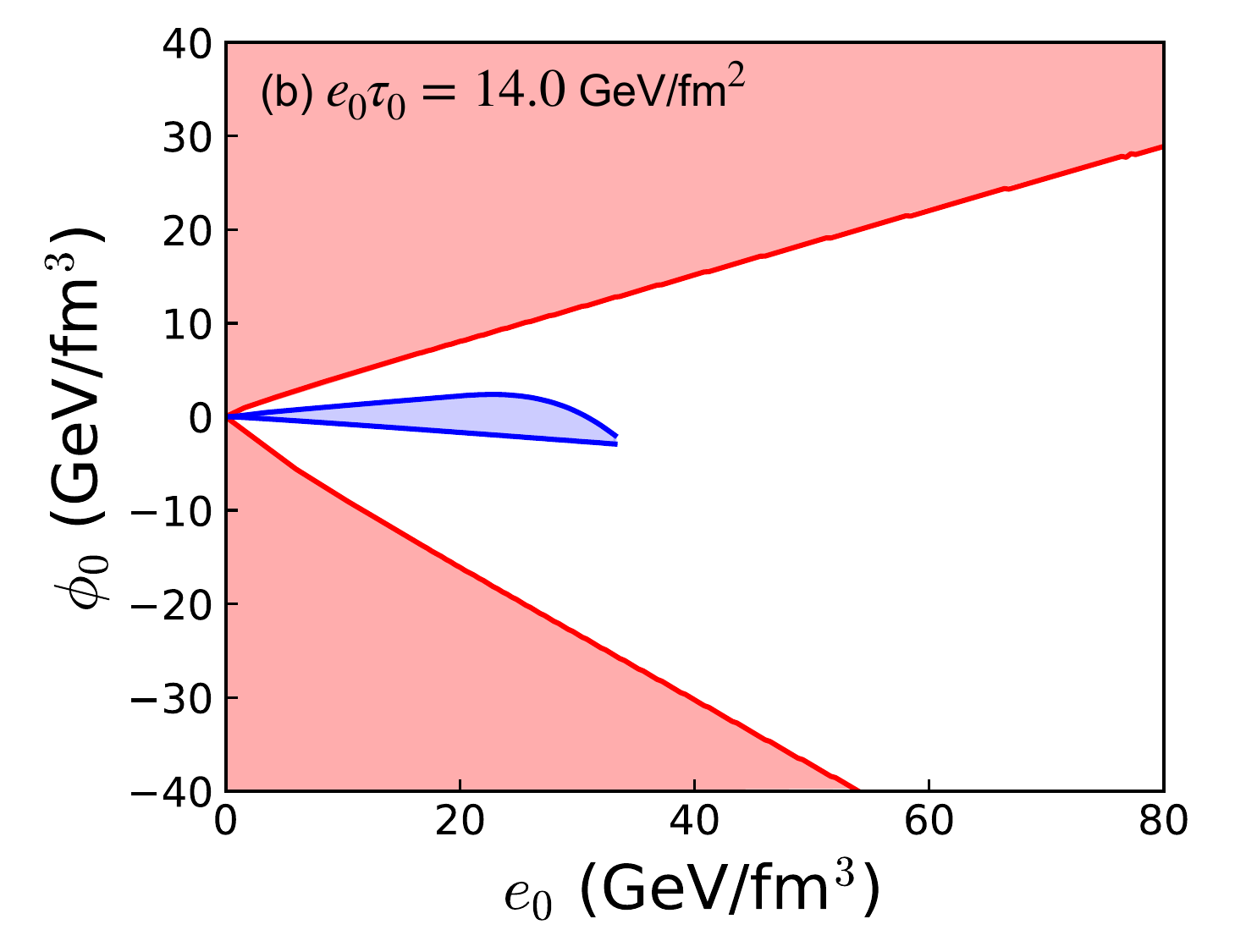}
    \caption{
    The same as Fig.~\ref{fig:result4} but employing the lattice EoS.
    (a) $e_0 \tau_0 = 5$ GeV/fm$^2$ at RHIC and (b) $14$ GeV/fm$^2$ at LHC energy.
     }
    \label{fig:result7}
\end{figure}

One finds in Fig.~\ref{fig:result7} that maximum initial energy density and minimum initial proper time are $e_{0,\mathrm{max}} \approx 11$ GeV/fm$^3$ and $\tau_{0,\mathrm{min}} \approx 0.5$ fm in top 5\% Au+Au collisions at $\sqrt{s_{NN}} = 200$ GeV and $e_{0,\mathrm{max}} \approx 35$ GeV/fm$^3$ and $\tau_{0,\mathrm{min}} \approx 0.4$ fm in top 5\% Pb+Pb collisions at $\sqrt{s_{NN}}=2.76$ TeV.
In comparison with the results from the conformal EoS, it turns out that the acceptable regions are broadened in the lattice EoS case.

Note that one cannot obtain a simple formula to connect the Bjorken energy density with the minimum initial proper time in the lattice EoS case  since it depends on the energy or temperature scale. 

\section{Summary}
\label{sec:summary}

We analyzed how far the one-dimensional expanding system can be away from local thermal equilibrium and constrain the regions of acceptable initial conditions from the nonlinear causality. 

We first obtained the necessary and the sufficient conditions of nonlinear causality in one-dimensional expanding systems which obey the BRSSS equation.
The necessary and the sufficient conditions are stated as nine and eight inequalities, respectively, among the thermodynamic variables, the shear pressure, and the transport coefficients.
Using the conformal EoS and a particular set of transport coefficients, we obtained $-0.47 < \phi/(e+p) < 0.23$ and $-0.07 < \phi/(e+p) < 0.07$ from the necessary and the sufficient conditions, respectively.
From these results, we concluded that the inverse Reynolds number must be small, meaning that the system must be close to the local equilibrium, in order to obey the relativistic causality.

Given initial conditions for the energy density and the shear pressure, we next investigated whether the solution violated or obeyed these conditions during the time evolution.
By monitoring the solution of the equations of motion, we checked whether the necessary and the sufficient conditions were respected at each time step.
We found that, due to the large expansion rate at the initial stage of relativistic heavy-ion collisions, causality can be violated during a course of evolution even if the system starts from a local equilibrium state.

With additional constraints from measured transverse energy per rapidity at RHIC and LHC, we further constrained the regions of acausal and causal initial conditions.
We found the acceptable initial conditions for the systems guaranteed to be causal are highly constrained from a combination of nonlinear causality and the measured transverse energy.
We derived a simple formula to quantify minimum initial time as $\tau_{0,\mathrm{min}}\enskip (\mathrm{fm}) \approx 1.6 \times [(1/S)dE_{T}/dy\enskip (\mathrm{GeV/fm}^{2})]^{-1/3}$ for an $N_f=3$ conformal system.
Using this formula, maximum initial energy density and minimum initial proper time are obtained as $e_{0,\mathrm{max}} \approx 5$ GeV/fm$^3$ and $\tau_{0,\mathrm{min}} \approx 1$ fm in top 5\% Au+Au collisions at $\sqrt{s_{NN}} = 200$ GeV  and $e_{0,\mathrm{max}} \approx 20$ GeV/fm$^3$ and $\tau_{0,\mathrm{min}} \approx 0.7$ fm   in top 5\% Pb+Pb collisions at $\sqrt{s_{NN}}=2.76$ TeV  in the conformal limit.
Although these constraints were obtained from  the sufficient conditions of nonlinear causality, these could be relaxed if one  chooses only the necessary conditions.

By employing the lattice EoS, we refined the results obtained using the conformal EoS.
Maximum initial energy density and minimum initial proper time are obtained as $e_{0,\mathrm{max}} \approx 11$ GeV/fm$^3$ and $\tau_{0,\mathrm{min}} \approx 0.5$ fm in top 5\% Au+Au collisions at $\sqrt{s_{NN}} = 200$ GeV  and $e_{0,\mathrm{max}} \approx 35$ GeV/fm$^3$ and $\tau_{0,\mathrm{min}} \approx 0.4$ fm  in top 5\% Pb+Pb collisions at $\sqrt{s_{NN}}=2.76$ TeV.
In comparison with the results with the conformal EoS case, a bigger $e_{0,\mathrm{max}}$ and a smaller  $\tau_{0,\mathrm{min}}$ are acceptable in the case of lattice EoS.

In this paper, we employed the BRSSS equation as a constitutive equation to demonstrate how the inverse Reynolds number is related to violation of causality.
Since the constitutive relation can be also obtained in various approaches, varieties of the second order constitutive equations can be found in the literature (e.g., Refs.~\cite{Hiscock:1983zz,Baier:2006um, Betz:2008me,Betz:2009zz,Denicol:2012cn,Monnai:2010qp,Tsumura:2006hnr,Tsumura:2009vm,Tsumura:2012ss,Jaiswal:2013npa}).
This could give a theoretical systematic uncertainty for constraints on initial conditions. 
A detailed analysis by employing other constitutive equations as well as parameter dependence of causality conditions on transport coefficients will be made in future analysis.

We emphasize here that we are not arguing against the existence of attractor behavior in this paper. 
Rather, we claim that a framework other than the M\"uller--Israel--Stewart type constitutive equations in relativistic dissipative dynamics is indispensable toward comprehensive understanding of hydrodynamization and/or  thermalization in relativistic heavy-ion collisions.
A microscopic description based on the kinetic theory might be such a framework \cite{Heller:2016rtz, Blaizot:2017ucy, Romatschke:2017vte, Behtash:2017wqg,Heller:2018qvh,Strickland:2018ayk,Denicol:2019lio,Heller:2020anv,Kamata:2020mka,Almaalol:2020rnu,Chattopadhyay:2021ive,Jaiswal:2022udf,Brewer:2022ifw}.
Just as we have examined the conditions for the applicability of relativistic hydrodynamics through nonlinear causality in this paper, we need to carefully consider the applicability of the Boltzmann equation.
In particular, the density of the system must be sufficiently low so that the truncation of the Bogoliubov--Born--Green--Kirkwood--Yvon (BBGKY) hierarchy is justified \cite{LandauLifshitzPhysicalKinetics}.
This would be challenging in the very early stage of high-energy nuclear collisions in which the parton density is extremely high.

A phenomenological description of pre-hydrodynamic stage based on the core-corona picture \cite{Werner:2007bf,Pierog:2013ria,Werner:2013tya,Werner:2023jps,Kanakubo:2018vkl,Kanakubo:2019ogh,Kanakubo:2021qcw,Kanakubo:2022ual} could be useful.
The local system is considered to be a fluid (the core) when its density is high enough to be equilibrated and its dynamics is described by using relativistic hydrodynamics.
Then, when the violation of causality happens locally in hydrodynamic simulations, one regards such system as nonequilibrated matter (the corona) and switches its description from a macroscopic hydrodynamics to a microscopic kinetic theory. 
A systematic analysis from a dynamical model free from violation of causality would be indispensable towards more quantitative understanding of the quark gluon plasma created in relativistic heavy-ion  collisions.

The validity of numerical simulations of relativistic dissipative hydrodynamics has been scrutinized through the necessary and the sufficient conditions of nonlinear causality 
\cite{plumberg2022causality,Chiu:2021muk,daSilva:2022xwu,ExTrEMe:2023nhy,Domingues:2024pom} and it turns out that causality is indeed violated in the early stage and/or in peripheral regions.
The violation of causality is intimately related to instability (see, e.g., Ref.~\cite{Gavassino:2021owo}). 
Nevertheless, numerical simulations of relativistic dissipative hydrodynamics seemingly  work well without any instability behaviors.
However, it is by no means trivial whether instabilities arise when hydrodynamic fluctuations act as perturbations on a local equilibrium system.
This becomes an important issue in sophisticated (3+1)-dimensional numerical simulations  based on relativistic fluctuating hydrodynamics \cite{Murase:2015oie,Sakai:2020pjw,Sakai:2021rug,Kuroki:2023ebq}.

\section*{Acknowledgement}
The work by T.H. was partly supported by JSPS KAKENHI Grant No. JP23K03395.

\appendix

\section{One-dimensional expansion of a conformal system}
\label{sec:hydrodynamization}

In this Appendix \ref{sec:hydrodynamization}, we discuss ``time" evolution of a conformal system which undergoes one-dimensional boost invariant expansion following the discussion in Refs.~\cite{heller2015hydrodynamics,Florkowski:2017olj}.
In particular, we present gradient vectors of solutions to capture those solutions' global behavior.

We define the dimensionless variables, $w$ and $f$, as
\begin{align}
    w=\tau T, \quad f = \frac{3}{2}\tau \frac{1}{w} \frac{dw}{d\tau }.
\end{align}
Here $\tau$ and $T$ are the proper time and the temperature, respectively.
Note that our definition of a variable $f$ is a factor $3/2$  different from the original one  \cite{heller2015hydrodynamics} for later purposes.
By using the equation of motion (\ref{eq:Bjorken_eq_for_energy_density}) together with $e \propto T^4$ and $p=e/3$ in the relativistic conformal system, a variable $f$ can be also written by using the inverse Reynolds number (\ref{eq:inverse_Reynolds_no}) as
\begin{align}
     f(\tau)=1+\frac{3}{8}\frac{\phi(\tau)}{e(\tau)} = 1\pm \frac{1}{2}Re^{-1}(\tau).\label{eq:equilibrium_measure}
\end{align}
Here the sign in front of $Re^{-1}$ corresponds to the sign of $\phi$.
It should be noted here that effective ``temperature" can be defined from the energy density in the conformal system even when the system is far away from the local equilibrium. 

By inserting the boost invariant solution \cite{Bjorken:1982qr}, $u^{\mu}_{\mathrm{Bj}} =(t,0,0,z)/\tau$,  into the constitutive equation from gradient expansion (\ref{eq:BRSSS}) and neglecting the last term in the right hand side which corresponds to a quadratic term of $\pi^{\mu \nu}$ in Eq.~(\ref{eq:BRSSS_resum}), one obtains
\begin{align}
    \phi_{\mathrm{ge}}=\frac{4}{3}\frac{ \eta}{\tau} + \frac{8}{9}\frac{\eta \tau_{\pi}}{\tau^2}.
\end{align}
Therefore, $f(w)$ is determined in the gradient expansion without solving the equation of motion as
\begin{align}
    f_{\mathrm{ge}}(w) &= 1+\frac{3}{8}\frac{\phi_{\mathrm{ge}}}{e} = 1+\frac{1}{2}\frac{\phi_{\mathrm{ge}}}{sT} \nonumber \\
    & = 1 +\frac{2}{3}C_\eta w^{-1} + \frac{4}{9}C_{\tau_\pi}C_{\eta}w^{-2},\label{eq:f_w_grad_exp}\\
    Re^{-1} & = \frac{4}{3}C_\eta w^{-1} + \frac{8}{9}C_{\tau_\pi}C_\eta w^{-2}. \label{eq:Re-1_grad_exp}
\end{align}

To interpret the meaning of $w$ and $f$, we suppose the ideal fluid case ($\phi = Re^{-1} = 0$ and $T\propto \tau^{-1/3}$).
Then these variables are
\begin{align}
     w \propto \tau ^{\frac{2}{3}}, \quad f=1.
\end{align}
Thus $w$ can be regard as a ``time" variable of the conformal system\footnote{When the inverse Reynolds number is moderate, the time dependence of $w$ must be slightly modified. Even in this case, it could be still regarded as a time variable. However, when the inverse Reynolds number is extremely large, interpretation of the variable $w$ as a time is no longer available since $w$ decreases with the proper time $\tau$.}  and $f$ can be called ``equilibrium measure" because deviation of $f$ from unity directly quantifies the half of the inverse Reynolds number.  
By using these variables, a set of simultaneous differential equations (\ref{eq:Bjorken_eq_for_energy_density}) and (\ref{eq:Bjorken_eq_for_constitutive}) with the conformal EoS (\ref{eq:conformal_eos}) is reduced to a single ordinary differential equation,
\begin{align}
     C_{\tau_\pi} &w f \frac{df}{dw} +4C_{\tau_\pi}(f-1)^2-C_\eta\nonumber\\
     &+\frac{3}{2}\frac{C_{\lambda_1}}{C_\eta}w(f-1)^2+\frac{3}{2}w(f-1)=0. \label{eq:f-w}
\end{align}
Note that the transport coefficients in Eqs.~(\ref{eq:transport_coeff_confomal}) and (\ref{eq:transport_coeff_AdSCFT}) are employed here.
Equation (\ref{eq:f-w})  can be regarded also as a ``time" evolution of the inverse Reynolds number and is also rewritten as
\begin{align}
    \frac{1}{2}C_{\tau_\pi}w \left(1+\frac{1}{2}Re^{-1} \right)&\frac{d Re^{-1}}{dw} + C_{\tau_\pi}(Re^{-1})^2 - C_\eta \nonumber \\
    &+ \frac{3}{8}\frac{C_{\lambda_1}}{C_\eta}w(Re^{-1})^2 + \frac{3}{4}(Re^{-1}) = 0.
\end{align}

Equation (\ref{eq:f_w_grad_exp}) can be also obtained by neglecting a nonlinear term in (\ref{eq:f-w}), expanding a solution into the series of the inverse of $w$ as $f(w) = \sum_{n=0}^{\infty}r_n w^{-n}$, and taking the terms up to $n=2$.

\begin{figure}[htpb]
    \centering
    \includegraphics[clip,width=\linewidth]{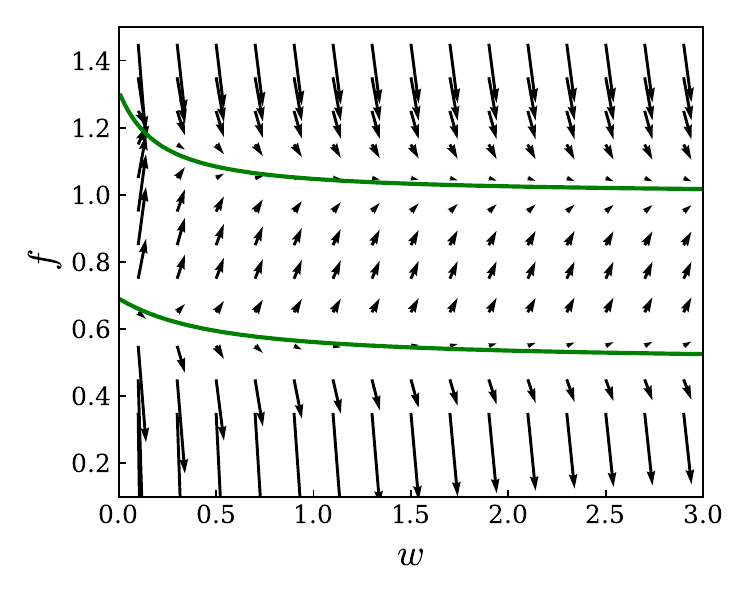}
    \caption{
    Gradient of solutions for Eq.~(\ref{eq:f-w}) at various points in the $w$--$f$ plane.
    Each arrow shows a vector ($\Delta w$, $\Delta f$) and the upper and the lower curves show the solution of $df/dw =0$, for $f=f_0^{+}(w)$ and $f=f_0^{-}(w)$, respectively.
     }
    \label{fig:flow_diagram}
\end{figure}

To capture the global behavior of the solutions, it would be helpful to draw a ``flow" diagram from the equation of motion (\ref{eq:f-w}). 
By inserting $df/dw=0$ into Eq.~(\ref{eq:f-w}), one obtains a quadratic equation with respect to $f$.
Its solutions become
\begin{align}
    f_0^{\pm}(w) = 1 + \frac{-3w\pm \sqrt{9w^2+64C_{\tau_\pi}C_\eta + 24C_{\lambda_1}w}}{16C_{\tau_\pi}+6\frac{C_{\lambda_1}}{C_\eta}w},
\end{align}
where $f_0^\pm(w \rightarrow 0) = 1 \pm \frac{1}{2}\sqrt{\frac{C_\eta}{C_{\tau_\pi}}}$.
Figure \ref{fig:flow_diagram} shows ``flow" vectors $(\Delta w, \Delta f)$ and $f=f_0^{\pm}(w)$ in the $w$-$f$ plane. 
From Fig.~\ref{fig:flow_diagram}, $df/dw>0$ in the region between $f=f_0^{+}(w)$ and $f=f_0^{-}(w)$, while $df/dw<0$ in the others.
The gradient of a solution changes its sign when it hits $f=f_0^{+}(w)$. It should be noted that $f=f_0^{+}(w)$ is not the hydrodynamic attractor solution but merely the boundary of changing the sign of $df/dw$.
The other solution, $f=f_0^{-}(w)$, is  ``unstable" in the sense that the solutions, $f(w)$, deviate from $f=f_0^{-}(w)$ due to $df/dw>0$ ($df/dw<0$) above (below) this boundary.
It is often stated that any initial conditions converge to the so-called hydrodynamic attractor in this system. However, if the system starts from an initial condition below $f=f_0^{-}(w)$, the solution no longer converges to the hydrodynamic attractor.

\section{Causality conditions}
\label{sec:causality appendix}
In this appendix \ref{sec:causality appendix}, we briefly review the necessary and the sufficient conditions of nonlinear causality in relativistic hydrodynamics following Ref.~\cite{bemfica2021nonlinear} and  then discuss a relation between nonlinear causality  and the dissipative quantities in one-dimensional expanding system.
In particular, we exhibit how the characteristic velocity is related to the equilibrium measure and, in turn, the inverse Reynolds number (\ref{eq:inverse_Reynolds_no}).

Relativistic hydrodynamic equations (\ref{eq:energy-momentum-conservation}) with the BRSSS constitutive equation (\ref{eq:BRSSS_resum}) \cite{BRSSS}  are summarized as 
\begin{align}
   & u^\alpha \partial_\alpha e + (e+p)\partial_\alpha u^\alpha = \pi^{\mu\nu} \nabla_{\langle\mu} u_{\nu\rangle}, \label{eq:he1}\\
   & (e+p)u^\beta \partial_\beta u_\alpha - c_s^2\Delta_\alpha^{\enskip \beta} \partial_\beta e  = - \Delta_\alpha^{\enskip \beta} \partial_\mu \pi^\mu_\beta,\label{eq:he2} \\
   & \tau_\pi \Delta^{\mu\nu}_{\enskip\alpha\beta}D\pi^{\alpha\beta} + \pi^{\mu\nu} = 2\eta \sigma^{\mu\nu} - \frac{4}{3}\tau_\pi \pi^{\mu\nu}\partial_\alpha u^\alpha. \label{eq:he3}
\end{align}
Equations (\ref{eq:he1})-(\ref{eq:he3}) can be compiled into a quasi-linear differential equation as
\begin{align}
    A^\alpha \partial_\alpha \Psi = F(\Psi),
\end{align}
where $\Psi = (e, u^\gamma, \pi^{0\gamma}, \pi^{1\gamma}, \pi^{2\gamma}, \pi^{3\gamma})^\mathrm{T}$ is a 21-dimensional column vector of unknown functions and $F(\Psi)$ is also a 21-dimensional column vector that does not contain any derivatives with respect to space-time coordinates.
A Lorentz vector $A^\alpha$ is a 21-by-21 matrix,
\begin{align}
A^\alpha = 
\begin{pmatrix}
    u^\alpha & (e+p)\delta^\alpha_{\enskip \gamma} - \pi^{\alpha}_{\enskip\gamma}& 0\\
    -c_s^2\Delta^\alpha_{\enskip \beta} & (e+p)u^\alpha g_{\gamma\beta} + \pi^\alpha_{\enskip \gamma} u_\beta & \delta^\alpha_{\enskip \sigma} g_{\gamma\beta}\\
    0 & \frac{4}{3}\tau_\pi \pi^{\mu\nu}\delta^\alpha_{\enskip\gamma} - 2\eta\Delta^{\mu\nu\alpha}_{\quad\gamma} & \tau_\pi \Delta^{\mu\nu}_{\enskip\sigma\gamma} u^\alpha 
\end{pmatrix}.
\end{align}

The characteristic surface $\Phi(x)$ is determined from the principal part, i.e. the highest order differential terms of the equations, by solving the characteristic equation,
\begin{align}
    \mathrm{det}(A^\alpha \xi_\alpha) = 0, \label{eq:characteristic equation}
\end{align}
where $\xi_\alpha = \partial_\alpha \Phi$. 

Causality conditions can be stated as follows: For the characteristic surface $\Phi(x)$ to be time-like or light-like, indicating that no information is propagated superluminally, 
its normal vector $\xi_\alpha$ needs to be spacelike or lightlike.

The gradient of the characteristic surface, $\xi_\alpha$, is decomposed into the time-like vector and space-like vector
\begin{align}
    \xi_\alpha = bu_\alpha + a_\alpha,
\end{align}
where $u^\alpha$ is a normalized timelike vector ($u_\alpha u^\alpha = 1$).
When $\xi_\alpha$ is a spacelike or light-like vector,
\begin{align}
     \xi^\alpha \xi_\alpha = b^2 + a \cdot a \le 0.
\end{align}
We define a new variable
\begin{align}
    v_{c}^2 =-\frac{b^2}{a \cdot a} \equiv k, \label{eq:characteristic_velocity}
\end{align}
and rewrite the condition as
\begin{align}
    0 \le k \le 1. \label{eq: causality condition}
\end{align}
This variable $v_c$ looks like a velocity, hence one may call it as a characteristic velocity.
In fact, this represents a response velocity of the system locally against the external perturbation.

Since the shear stress tensor $\pi^{\mu \nu}$ is a symmetric and traceless tensor, one obtains the four eigenvectors $e_{I}^\mu$ and their eigenvalues $\Lambda_{I}$  ($I= 0, 1, 2, 3$) from the eigenvalue equation, 
\begin{align}
    \pi^{\mu}_{\enskip \nu}e_{I}^\nu  = \Lambda_{I}e_{I}^\nu.
\end{align}
From the transverse condition $\pi^\mu_{\enskip \nu}u^\nu = 0$,  $u^\mu (\equiv e_{I=0}^\mu)$ is a trivial eigenvector of the shear stress tensor with its eigenvalue $\Lambda_0 = 0$.
The traceless condition $\pi^\mu_{\enskip \mu}=0$ leads to 
\begin{align}
 \sum_{I=0}^{3}\Lambda_I = \sum_{i=1}^{3}\Lambda_i = 0.
\end{align}
Thus one can choose the eigenvalues as $\Lambda_{1}\le \Lambda_{2} \le \Lambda_{3}$, $\Lambda_{1}<0$, and $\Lambda_{3}>0$.
Since $e_{I}^\mu$ ($I=0, 1, 2, 3$) is a complete set in $\mathbb{R}^4$, one can regard this as a tetrad which obeys $g_{\mu \nu} e_{I}^\mu e_{J}^\nu= g_{IJ}$, where $g_{IJ} = \mathrm{diag}(1, -1, -1, -1)$.

By using the completeness relation,
\begin{align}
    \delta^{\mu}_{\enskip \nu} = \sum_{I=0}^{3}e_{I}^{\mu} e^{I}_{\nu} = u^\mu u_\nu + \sum_{i=1}^{3} e_{i}^\mu e^{i}_\nu,
\end{align}
one can expand an arbitrary Lorentz vector. 
The gradient of the surface vector can be also expanded as
\begin{align}
    \xi^\mu &= bu^\mu + \sum_{i=1}^{3} a^{i} e_{i}^\mu, \\
    \xi^{I=0} &= b, \enskip \xi^{i} = a^{i}, \enskip (i= 1, 2, 3)
\end{align}
where $a_{I} = e_{I}^\mu a_\mu$. Thus one finds 
\begin{align}
    a_{I=0} = e_{I=0}^\mu a_\mu = u^\mu a_\mu = 0.
\end{align}
Therefore the spacelike vector $a^\mu$ obeys
\begin{align}
    a \cdot a = a^\mu a_\mu &= a^{I}a_{I} = a_{I=0}^2 - \sum_{i=1}^{3} a_{i}^{2} =- \sum_{i=1}^{3} a_{i}^{2}<0,\\
   &\hat{a}_i = \frac{a_i}{\sqrt{-a\cdot a}}, \quad
\sum_{i=1}^{3}\hat{a}_{i}^2 = 1. 
\end{align}

One obtains from Eq.~(\ref{eq:characteristic equation})
\begin{align}
     &\det(A^\alpha\xi_\alpha) 
    = b^{13}\tau_\pi^{16}\mathrm{det}(M) = 0.
\end{align}
Thus, besides 13 trivial solutions $b=0$ and consequently $v_c =0$ from Eq.~(\ref{eq:characteristic_velocity}), the equation we have to solve to obtain the characteristic surface is
\begin{align}
    &\det(M)(k) =  (a\cdot a)^4 m_0(k) m_1(k) m_2(k) m_3(k) G(k) =0, \label{eq:determinant_M}
\end{align}
where
\begin{align}
m_I(k) &= (e+p+\Lambda_I)k  - \frac{\eta}{\tau_\pi},\enskip (I=0, 1, 2, 3) \\
    G(k) &=  1-\sum_{i=1}^{3} \frac{\left[\frac{\eta}{3\tau_\pi} + \frac{4}{3} \Lambda_i + (e+p + \Lambda_i)c_s^2 \right]\hat{a}_i^2}{m_i(k)} \nonumber\\
    & - \frac{4}{3}c_s^2 \sum_{i\neq j=1}^{3}\frac{(\Lambda_i -\Lambda_j)^2\hat{a}_i^2\hat{a}_j^2}{m_i(k) m_j(k)}.\label{eq:Gk}
\end{align}

We require all characteristic velocities to be causal as the necessary conditions, not as the necessary and sufficient conditions. Hence even an extreme choice of the gradient of characteristic surface as above would be satisfied as the necessary conditions.
Therefore we may choose $\hat{a}^2_i = 1$ for $i = 1$ or $2$ or $3$ and $\hat{a}^2_j = 0$ for $i \neq j$ \cite{bemfica2021nonlinear}.
Then $G(k)$ in Eq.~(\ref{eq:Gk}) becomes
\begin{align}
        G(k; i) &=  1- \frac{\frac{\eta}{3\tau_\pi} + \frac{4}{3} \Lambda_i + (e+p + \Lambda_i)c_s^2 }{m_i(k)}. \enskip  (i = 1, 2, 3) \label{eq:specific G(k)}
\end{align}

Since the number of solutions for Eq.~(\ref{eq:characteristic equation}) is 21 and one has already 13 trivial solutions ($b=0$), Eq.~(\ref{eq:determinant_M}) has four independent solutions each of which has two solutions of characteristic velocity.
Let $k_i$ be the solutions of the characteristic equation (\ref{eq:determinant_M}), these solutions satisfy $m_0(k_0) = 0$ and
\begin{align}
  & m_{2}(k_{2})= 0, \enskip m_{3}(k_{3}) = 0, \enskip m_1(k_{4})G(k_{4}; i=1)   = 0, \\
&\quad \mathrm{or} \nonumber \\
  & m_{3}(k_{3})= 0, \enskip m_{1}(k_{1})= 0, \enskip m_2(k_{5})G(k_{5}; i=2)   = 0,\\
&\quad  \mathrm{or} \nonumber \\
  &  m_{1}(k_{1}) = 0, \enskip m_{2}(k_{2})= 0, \enskip m_3(k_{6})G(k_{6}; i=3)   = 0. 
\end{align}
It should be noted that three different sets of four independent solutions above correspond to a specific choice of $\hat{a}^2_i = 1$ ($i = 1$ or $2$ or $3$) to obtain  Eq.~(\ref{eq:specific G(k)}).

Necessary conditions for the system to be causal indicate 
that all the square of the characteristic velocities satisfy Eq.~(\ref{eq: causality condition}), namely, $0\leq k_i \leq 1$ \enskip ($i=0, 1, 2, \cdots, 6$):
\begin{align}
    &0 \leq k_i =\frac{\eta }{\tau_\pi(e+p+\Lambda_i)}\leq 1, \enskip (i=0, 1, 2, 3)\label{eq:characteristic velocity1} \\
    & 0 \leq k_i =  c_s^2+\frac{4}{3}\frac{\left( \frac{\eta}{\tau_\pi}  + \Lambda_{i-3} \right) }{e+p+\Lambda_{i-3}}  \leq 1. \enskip (i=4, 5, 6) \label{eq:characteristic velocity2}
\end{align}
Note  that  $k_1$ in Eq.~(\ref{eq:characteristic velocity1}) gives the most severe condition for $k \le 1$ among $i = 0$, $1$, $2$, and $3$ due to $\Lambda_1< 0 < \Lambda_3$.

In the case of a one-dimensional expanding system, shear  pressures become $\Lambda_0 = 0$ and $\Lambda_i (i=1, 2, 3) = -\phi/2$  or $\phi$.
Finally one obtains Eqs.~(\ref{eq:Necessary_BRSSS1})-(\ref{eq:Necessary_BRSSS6}) as necessary conditions of nonlinear causality for one-dimensional expanding system which obeys the BRSSS equation.

Before discussing the sufficient conditions, we define a polynomial function $g(k)$ from Eq.~(\ref{eq:determinant_M}) as
\begin{align}
    g(k) &= m_1(k)m_2(k)m_3(k)G(k).
\end{align}
Since $g(k)$ is the polynomial function at the third order, one factorizes $g(k)$ as 
\begin{align}
g(k) &= \left[\prod_{i=1}^3(e+p+\Lambda_i)\right](k-\tilde{k}_1)(k-\tilde{k}_2)(k-\tilde{k}_3),
\end{align}
where $\tilde{k}_i$ ($i=1, 2, 3$) are solutions of $g(k)=0$.
Note that $\tilde{k}_i$ can be different from $k_i$ since we supposed a specific form of $G(k)$ as in Eq.~(\ref{eq:specific G(k)}) to obtain the characteristic velocity, $k_i$, in the necessary conditions.

Since the solutions are supposed to obey an inequality (\ref{eq: causality condition}), namely $0 \le \tilde{k}_i \le 1$, and the sign in front of $k^3$ is also supposed to be positive, the sufficient conditions for the system to be causal indicate that $g(k)$ should satisfy 
\begin{align}
    g(k<0)<0,\\
    g(k>1)>0.
\end{align}

In the case of a one-dimensional expanding system  which obeys the BRSSS equation, one obtains Eqs.~(\ref{eq:suffficient_positive_phi_first})-(\ref{eq:suffficient_negative_phi_last}) as sufficient conditions of nonlinear causality.

\begin{figure}[htpb]
    \centering
    \includegraphics[clip,width=\linewidth]{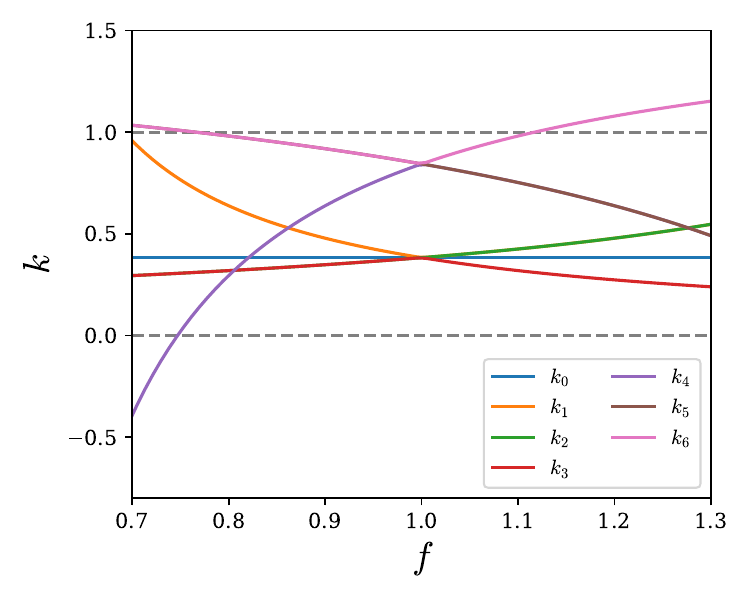}
    \caption{
    Squares of characteristic velocity  (\ref{eq:characteristic velocity1}) and (\ref{eq:characteristic velocity2}) as functions of equilibrium measure (\ref{eq:equilibrium_measure}) in one-dimensional expanding conformal system.
    Seven different forms of squares of characteristic velocity, $k_i$ ($i=0, \cdots, 6$),  are shown for comparison.
     }
    \label{fig:ki_vs_f}
\end{figure}

Equilibrium measure dependence of squares of characteristic velocity, (\ref{eq:characteristic velocity1}) and (\ref{eq:characteristic velocity2}), in the case of one-dimensional expanding system is shown in Fig.~\ref{fig:ki_vs_f}. 
When the system is in local equilibrium, namely $f=1$, the squares of characteristic velocity have $k_i \approx 0.38$ for $i=0$, $1$, $2$, and $3$ and $k_i \approx 0.84$ for $i=4$, $5$, and $6$.
Hence the system obeys the necessary conditions of nonlinear causality in local equilibrium. 
When $f$ increases or decreases with increasing inverse Reynolds number, one of the solutions of the characteristic equations, $k_6$, first violates the necessary conditions.
This provides the inequality of the inverse Reynolds number as shown in Eq.~(\ref{eq:Necessary_Bj}).

\bibliography{reference}

\begin{thebibliography}{72}%
\makeatletter
\providecommand \@ifxundefined [1]{%
 \@ifx{#1\undefined}
}%
\providecommand \@ifnum [1]{%
 \ifnum #1\expandafter \@firstoftwo
 \else \expandafter \@secondoftwo
 \fi
}%
\providecommand \@ifx [1]{%
 \ifx #1\expandafter \@firstoftwo
 \else \expandafter \@secondoftwo
 \fi
}%
\providecommand \natexlab [1]{#1}%
\providecommand \enquote  [1]{``#1''}%
\providecommand \bibnamefont  [1]{#1}%
\providecommand \bibfnamefont [1]{#1}%
\providecommand \citenamefont [1]{#1}%
\providecommand \href@noop [0]{\@secondoftwo}%
\providecommand \href [0]{\begingroup \@sanitize@url \@href}%
\providecommand \@href[1]{\@@startlink{#1}\@@href}%
\providecommand \@@href[1]{\endgroup#1\@@endlink}%
\providecommand \@sanitize@url [0]{\catcode `\\12\catcode `\$12\catcode
  `\&12\catcode `\#12\catcode `\^12\catcode `\_12\catcode `\%12\relax}%
\providecommand \@@startlink[1]{}%
\providecommand \@@endlink[0]{}%
\providecommand \url  [0]{\begingroup\@sanitize@url \@url }%
\providecommand \@url [1]{\endgroup\@href {#1}{\urlprefix }}%
\providecommand \urlprefix  [0]{URL }%
\providecommand \Eprint [0]{\href }%
\providecommand \doibase [0]{http://dx.doi.org/}%
\providecommand \selectlanguage [0]{\@gobble}%
\providecommand \bibinfo  [0]{\@secondoftwo}%
\providecommand \bibfield  [0]{\@secondoftwo}%
\providecommand \translation [1]{[#1]}%
\providecommand \BibitemOpen [0]{}%
\providecommand \bibitemStop [0]{}%
\providecommand \bibitemNoStop [0]{.\EOS\space}%
\providecommand \EOS [0]{\spacefactor3000\relax}%
\providecommand \BibitemShut  [1]{\csname bibitem#1\endcsname}%
\let\auto@bib@innerbib\@empty
\bibitem [{\citenamefont {Yagi}\ \emph {et~al.}(2005)\citenamefont {Yagi},
  \citenamefont {Hatsuda},\ and\ \citenamefont {Miake}}]{Yagi:2005yb}%
  \BibitemOpen
  \bibfield  {author} {\bibinfo {author} {\bibfnamefont {K.}~\bibnamefont
  {Yagi}}, \bibinfo {author} {\bibfnamefont {T.}~\bibnamefont {Hatsuda}}, \
  and\ \bibinfo {author} {\bibfnamefont {Y.}~\bibnamefont {Miake}},\
  }\href@noop {} {\emph {\bibinfo {title} {{Quark-Gluon Plasma}}}},\
  Vol.~\bibinfo {volume} {23}\ (\bibinfo  {publisher} {Cambridge University
  Press},\ \bibinfo {year} {2005})\BibitemShut {NoStop}%
\bibitem [{\citenamefont {Hirano}\ and\ \citenamefont
  {Gyulassy}(2006)}]{Hirano:2005wx}%
  \BibitemOpen
  \bibfield  {author} {\bibinfo {author} {\bibfnamefont {T.}~\bibnamefont
  {Hirano}}\ and\ \bibinfo {author} {\bibfnamefont {M.}~\bibnamefont
  {Gyulassy}},\ }\href {\doibase 10.1016/j.nuclphysa.2006.02.005} {\bibfield
  {journal} {\bibinfo  {journal} {Nucl. Phys. A}\ }\textbf {\bibinfo {volume}
  {769}},\ \bibinfo {pages} {71} (\bibinfo {year} {2006})},\ \Eprint
  {http://arxiv.org/abs/nucl-th/0506049} {arXiv:nucl-th/0506049} \BibitemShut
  {NoStop}%
\bibitem [{\citenamefont {Bernhard}\ \emph {et~al.}(2016)\citenamefont
  {Bernhard}, \citenamefont {Moreland}, \citenamefont {Bass}, \citenamefont
  {Liu},\ and\ \citenamefont {Heinz}}]{Bernhard:2016tnd}%
  \BibitemOpen
  \bibfield  {author} {\bibinfo {author} {\bibfnamefont {J.~E.}\ \bibnamefont
  {Bernhard}}, \bibinfo {author} {\bibfnamefont {J.~S.}\ \bibnamefont
  {Moreland}}, \bibinfo {author} {\bibfnamefont {S.~A.}\ \bibnamefont {Bass}},
  \bibinfo {author} {\bibfnamefont {J.}~\bibnamefont {Liu}}, \ and\ \bibinfo
  {author} {\bibfnamefont {U.}~\bibnamefont {Heinz}},\ }\href {\doibase
  10.1103/PhysRevC.94.024907} {\bibfield  {journal} {\bibinfo  {journal} {Phys.
  Rev. C}\ }\textbf {\bibinfo {volume} {94}},\ \bibinfo {pages} {024907}
  (\bibinfo {year} {2016})},\ \Eprint {http://arxiv.org/abs/1605.03954}
  {arXiv:1605.03954 [nucl-th]} \BibitemShut {NoStop}%
\bibitem [{\citenamefont {Bernhard}\ \emph {et~al.}(2019)\citenamefont
  {Bernhard}, \citenamefont {Moreland},\ and\ \citenamefont
  {Bass}}]{Bernhard:2019bmu}%
  \BibitemOpen
  \bibfield  {author} {\bibinfo {author} {\bibfnamefont {J.~E.}\ \bibnamefont
  {Bernhard}}, \bibinfo {author} {\bibfnamefont {J.~S.}\ \bibnamefont
  {Moreland}}, \ and\ \bibinfo {author} {\bibfnamefont {S.~A.}\ \bibnamefont
  {Bass}},\ }\href {\doibase 10.1038/s41567-019-0611-8} {\bibfield  {journal}
  {\bibinfo  {journal} {Nature Phys.}\ }\textbf {\bibinfo {volume} {15}},\
  \bibinfo {pages} {1113} (\bibinfo {year} {2019})}\BibitemShut {NoStop}%
\bibitem [{\citenamefont {Auvinen}\ \emph {et~al.}(2020)\citenamefont
  {Auvinen}, \citenamefont {Eskola}, \citenamefont {Huovinen}, \citenamefont
  {Niemi}, \citenamefont {Paatelainen},\ and\ \citenamefont
  {Petreczky}}]{Auvinen:2020mpc}%
  \BibitemOpen
  \bibfield  {author} {\bibinfo {author} {\bibfnamefont {J.}~\bibnamefont
  {Auvinen}}, \bibinfo {author} {\bibfnamefont {K.~J.}\ \bibnamefont {Eskola}},
  \bibinfo {author} {\bibfnamefont {P.}~\bibnamefont {Huovinen}}, \bibinfo
  {author} {\bibfnamefont {H.}~\bibnamefont {Niemi}}, \bibinfo {author}
  {\bibfnamefont {R.}~\bibnamefont {Paatelainen}}, \ and\ \bibinfo {author}
  {\bibfnamefont {P.}~\bibnamefont {Petreczky}},\ }\href {\doibase
  10.1103/PhysRevC.102.044911} {\bibfield  {journal} {\bibinfo  {journal}
  {Phys. Rev. C}\ }\textbf {\bibinfo {volume} {102}},\ \bibinfo {pages}
  {044911} (\bibinfo {year} {2020})},\ \Eprint
  {http://arxiv.org/abs/2006.12499} {arXiv:2006.12499 [nucl-th]} \BibitemShut
  {NoStop}%
\bibitem [{\citenamefont {Nijs}\ \emph
  {et~al.}(2021{\natexlab{a}})\citenamefont {Nijs}, \citenamefont {van~der
  Schee}, \citenamefont {G\"ursoy},\ and\ \citenamefont
  {Snellings}}]{Nijs:2020ors}%
  \BibitemOpen
  \bibfield  {author} {\bibinfo {author} {\bibfnamefont {G.}~\bibnamefont
  {Nijs}}, \bibinfo {author} {\bibfnamefont {W.}~\bibnamefont {van~der Schee}},
  \bibinfo {author} {\bibfnamefont {U.}~\bibnamefont {G\"ursoy}}, \ and\
  \bibinfo {author} {\bibfnamefont {R.}~\bibnamefont {Snellings}},\ }\href
  {\doibase 10.1103/PhysRevLett.126.202301} {\bibfield  {journal} {\bibinfo
  {journal} {Phys. Rev. Lett.}\ }\textbf {\bibinfo {volume} {126}},\ \bibinfo
  {pages} {202301} (\bibinfo {year} {2021}{\natexlab{a}})},\ \Eprint
  {http://arxiv.org/abs/2010.15130} {arXiv:2010.15130 [nucl-th]} \BibitemShut
  {NoStop}%
\bibitem [{\citenamefont {Nijs}\ \emph
  {et~al.}(2021{\natexlab{b}})\citenamefont {Nijs}, \citenamefont {van~der
  Schee}, \citenamefont {G\"ursoy},\ and\ \citenamefont
  {Snellings}}]{Nijs:2020roc}%
  \BibitemOpen
  \bibfield  {author} {\bibinfo {author} {\bibfnamefont {G.}~\bibnamefont
  {Nijs}}, \bibinfo {author} {\bibfnamefont {W.}~\bibnamefont {van~der Schee}},
  \bibinfo {author} {\bibfnamefont {U.}~\bibnamefont {G\"ursoy}}, \ and\
  \bibinfo {author} {\bibfnamefont {R.}~\bibnamefont {Snellings}},\ }\href
  {\doibase 10.1103/PhysRevC.103.054909} {\bibfield  {journal} {\bibinfo
  {journal} {Phys. Rev. C}\ }\textbf {\bibinfo {volume} {103}},\ \bibinfo
  {pages} {054909} (\bibinfo {year} {2021}{\natexlab{b}})},\ \Eprint
  {http://arxiv.org/abs/2010.15134} {arXiv:2010.15134 [nucl-th]} \BibitemShut
  {NoStop}%
\bibitem [{\citenamefont {Parkkila}\ \emph {et~al.}(2021)\citenamefont
  {Parkkila}, \citenamefont {Onnerstad},\ and\ \citenamefont
  {Kim}}]{Parkkila:2021tqq}%
  \BibitemOpen
  \bibfield  {author} {\bibinfo {author} {\bibfnamefont {J.~E.}\ \bibnamefont
  {Parkkila}}, \bibinfo {author} {\bibfnamefont {A.}~\bibnamefont {Onnerstad}},
  \ and\ \bibinfo {author} {\bibfnamefont {D.~J.}\ \bibnamefont {Kim}},\ }\href
  {\doibase 10.1103/PhysRevC.104.054904} {\bibfield  {journal} {\bibinfo
  {journal} {Phys. Rev. C}\ }\textbf {\bibinfo {volume} {104}},\ \bibinfo
  {pages} {054904} (\bibinfo {year} {2021})},\ \Eprint
  {http://arxiv.org/abs/2106.05019} {arXiv:2106.05019 [hep-ph]} \BibitemShut
  {NoStop}%
\bibitem [{\citenamefont {Everett}\ \emph {et~al.}(2021)\citenamefont {Everett}
  \emph {et~al.}}]{JETSCAPE:2020mzn}%
  \BibitemOpen
  \bibfield  {author} {\bibinfo {author} {\bibfnamefont {D.}~\bibnamefont
  {Everett}} \emph {et~al.} (\bibinfo {collaboration} {JETSCAPE}),\ }\href
  {\doibase 10.1103/PhysRevC.103.054904} {\bibfield  {journal} {\bibinfo
  {journal} {Phys. Rev. C}\ }\textbf {\bibinfo {volume} {103}},\ \bibinfo
  {pages} {054904} (\bibinfo {year} {2021})},\ \Eprint
  {http://arxiv.org/abs/2011.01430} {arXiv:2011.01430 [hep-ph]} \BibitemShut
  {NoStop}%
\bibitem [{\citenamefont {Heffernan}\ \emph {et~al.}(2023)\citenamefont
  {Heffernan}, \citenamefont {Gale}, \citenamefont {Jeon},\ and\ \citenamefont
  {Paquet}}]{Heffernan:2023gye}%
  \BibitemOpen
  \bibfield  {author} {\bibinfo {author} {\bibfnamefont {M.~R.}\ \bibnamefont
  {Heffernan}}, \bibinfo {author} {\bibfnamefont {C.}~\bibnamefont {Gale}},
  \bibinfo {author} {\bibfnamefont {S.}~\bibnamefont {Jeon}}, \ and\ \bibinfo
  {author} {\bibfnamefont {J.-F.}\ \bibnamefont {Paquet}},\ }\href@noop {} {\
  (\bibinfo {year} {2023})},\ \Eprint {http://arxiv.org/abs/2306.09619}
  {arXiv:2306.09619 [nucl-th]} \BibitemShut {NoStop}%
\bibitem [{\citenamefont {Shen}\ \emph {et~al.}(2023)\citenamefont {Shen},
  \citenamefont {Schenke},\ and\ \citenamefont {Zhao}}]{Shen:2023awv}%
  \BibitemOpen
  \bibfield  {author} {\bibinfo {author} {\bibfnamefont {C.}~\bibnamefont
  {Shen}}, \bibinfo {author} {\bibfnamefont {B.}~\bibnamefont {Schenke}}, \
  and\ \bibinfo {author} {\bibfnamefont {W.}~\bibnamefont {Zhao}},\ }\href@noop
  {} {\  (\bibinfo {year} {2023})},\ \Eprint {http://arxiv.org/abs/2310.10787}
  {arXiv:2310.10787 [nucl-th]} \BibitemShut {NoStop}%
\bibitem [{\citenamefont {M{\"u}ller}(1967)}]{Muller:1967zza}%
  \BibitemOpen
  \bibfield  {author} {\bibinfo {author} {\bibfnamefont {I.}~\bibnamefont
  {M{\"u}ller}},\ }\href {\doibase 10.1007/BF01326412} {\bibfield  {journal}
  {\bibinfo  {journal} {Z. Phys.}\ }\textbf {\bibinfo {volume} {198}},\
  \bibinfo {pages} {329} (\bibinfo {year} {1967})}\BibitemShut {NoStop}%
\bibitem [{\citenamefont {Israel}(1976)}]{Israel:1976tn}%
  \BibitemOpen
  \bibfield  {author} {\bibinfo {author} {\bibfnamefont {W.}~\bibnamefont
  {Israel}},\ }\href {\doibase 10.1016/0003-4916(76)90064-6} {\bibfield
  {journal} {\bibinfo  {journal} {Annals Phys.}\ }\textbf {\bibinfo {volume}
  {100}},\ \bibinfo {pages} {310} (\bibinfo {year} {1976})}\BibitemShut
  {NoStop}%
\bibitem [{\citenamefont {Israel}\ and\ \citenamefont
  {Stewart}(1979)}]{Israel:1979wp}%
  \BibitemOpen
  \bibfield  {author} {\bibinfo {author} {\bibfnamefont {W.}~\bibnamefont
  {Israel}}\ and\ \bibinfo {author} {\bibfnamefont {J.~M.}\ \bibnamefont
  {Stewart}},\ }\href {\doibase 10.1016/0003-4916(79)90130-1} {\bibfield
  {journal} {\bibinfo  {journal} {Annals Phys.}\ }\textbf {\bibinfo {volume}
  {118}},\ \bibinfo {pages} {341} (\bibinfo {year} {1979})}\BibitemShut
  {NoStop}%
\bibitem [{\citenamefont {Hiscock}\ and\ \citenamefont
  {Lindblom}(1983)}]{Hiscock:1983zz}%
  \BibitemOpen
  \bibfield  {author} {\bibinfo {author} {\bibfnamefont {W.~A.}\ \bibnamefont
  {Hiscock}}\ and\ \bibinfo {author} {\bibfnamefont {L.}~\bibnamefont
  {Lindblom}},\ }\href {\doibase 10.1016/0003-4916(83)90288-9} {\bibfield
  {journal} {\bibinfo  {journal} {Annals Phys.}\ }\textbf {\bibinfo {volume}
  {151}},\ \bibinfo {pages} {466} (\bibinfo {year} {1983})}\BibitemShut
  {NoStop}%
\bibitem [{\citenamefont {Hiscock}\ and\ \citenamefont
  {Olson}(1989)}]{HISCOCK1989125}%
  \BibitemOpen
  \bibfield  {author} {\bibinfo {author} {\bibfnamefont {W.~A.}\ \bibnamefont
  {Hiscock}}\ and\ \bibinfo {author} {\bibfnamefont {T.~S.}\ \bibnamefont
  {Olson}},\ }\href {\doibase https://doi.org/10.1016/0375-9601(89)90772-X}
  {\bibfield  {journal} {\bibinfo  {journal} {Physics Letters A}\ }\textbf
  {\bibinfo {volume} {141}},\ \bibinfo {pages} {125} (\bibinfo {year}
  {1989})}\BibitemShut {NoStop}%
\bibitem [{\citenamefont {Denicol}\ \emph {et~al.}(2008)\citenamefont
  {Denicol}, \citenamefont {Kodama}, \citenamefont {Koide},\ and\ \citenamefont
  {Mota}}]{Denicol:2008ha}%
  \BibitemOpen
  \bibfield  {author} {\bibinfo {author} {\bibfnamefont {G.~S.}\ \bibnamefont
  {Denicol}}, \bibinfo {author} {\bibfnamefont {T.}~\bibnamefont {Kodama}},
  \bibinfo {author} {\bibfnamefont {T.}~\bibnamefont {Koide}}, \ and\ \bibinfo
  {author} {\bibfnamefont {P.}~\bibnamefont {Mota}},\ }\href {\doibase
  10.1088/0954-3899/35/11/115102} {\bibfield  {journal} {\bibinfo  {journal}
  {J. Phys. G}\ }\textbf {\bibinfo {volume} {35}},\ \bibinfo {pages} {115102}
  (\bibinfo {year} {2008})},\ \Eprint {http://arxiv.org/abs/0807.3120}
  {arXiv:0807.3120 [hep-ph]} \BibitemShut {NoStop}%
\bibitem [{\citenamefont {Pu}\ \emph {et~al.}(2010)\citenamefont {Pu},
  \citenamefont {Koide},\ and\ \citenamefont {Rischke}}]{Pu:2009fj}%
  \BibitemOpen
  \bibfield  {author} {\bibinfo {author} {\bibfnamefont {S.}~\bibnamefont
  {Pu}}, \bibinfo {author} {\bibfnamefont {T.}~\bibnamefont {Koide}}, \ and\
  \bibinfo {author} {\bibfnamefont {D.~H.}\ \bibnamefont {Rischke}},\ }\href
  {\doibase 10.1103/PhysRevD.81.114039} {\bibfield  {journal} {\bibinfo
  {journal} {Phys. Rev. D}\ }\textbf {\bibinfo {volume} {81}},\ \bibinfo
  {pages} {114039} (\bibinfo {year} {2010})},\ \Eprint
  {http://arxiv.org/abs/0907.3906} {arXiv:0907.3906 [hep-ph]} \BibitemShut
  {NoStop}%
\bibitem [{\citenamefont {Floerchinger}\ and\ \citenamefont
  {Grossi}(2018)}]{Floerchinger:2017cii}%
  \BibitemOpen
  \bibfield  {author} {\bibinfo {author} {\bibfnamefont {S.}~\bibnamefont
  {Floerchinger}}\ and\ \bibinfo {author} {\bibfnamefont {E.}~\bibnamefont
  {Grossi}},\ }\href {\doibase 10.1007/JHEP08(2018)186} {\bibfield  {journal}
  {\bibinfo  {journal} {JHEP}\ }\textbf {\bibinfo {volume} {08}},\ \bibinfo
  {pages} {186} (\bibinfo {year} {2018})},\ \Eprint
  {http://arxiv.org/abs/1711.06687} {arXiv:1711.06687 [nucl-th]} \BibitemShut
  {NoStop}%
\bibitem [{\citenamefont {Bemfica}\ \emph {et~al.}(2021)\citenamefont
  {Bemfica}, \citenamefont {Disconzi}, \citenamefont {Hoang}, \citenamefont
  {Noronha},\ and\ \citenamefont {Radosz}}]{bemfica2021nonlinear}%
  \BibitemOpen
  \bibfield  {author} {\bibinfo {author} {\bibfnamefont {F.~S.}\ \bibnamefont
  {Bemfica}}, \bibinfo {author} {\bibfnamefont {M.~M.}\ \bibnamefont
  {Disconzi}}, \bibinfo {author} {\bibfnamefont {V.}~\bibnamefont {Hoang}},
  \bibinfo {author} {\bibfnamefont {J.}~\bibnamefont {Noronha}}, \ and\
  \bibinfo {author} {\bibfnamefont {M.}~\bibnamefont {Radosz}},\ }\href
  {\doibase 10.1103/PhysRevLett.126.222301} {\bibfield  {journal} {\bibinfo
  {journal} {Phys. Rev. Lett.}\ }\textbf {\bibinfo {volume} {126}},\ \bibinfo
  {pages} {222301} (\bibinfo {year} {2021})},\ \Eprint
  {http://arxiv.org/abs/2005.11632} {arXiv:2005.11632 [hep-th]} \BibitemShut
  {NoStop}%
\bibitem [{\citenamefont {Plumberg}\ \emph {et~al.}(2022)\citenamefont
  {Plumberg}, \citenamefont {Almaalol}, \citenamefont {Dore}, \citenamefont
  {Noronha},\ and\ \citenamefont {Noronha-Hostler}}]{plumberg2022causality}%
  \BibitemOpen
  \bibfield  {author} {\bibinfo {author} {\bibfnamefont {C.}~\bibnamefont
  {Plumberg}}, \bibinfo {author} {\bibfnamefont {D.}~\bibnamefont {Almaalol}},
  \bibinfo {author} {\bibfnamefont {T.}~\bibnamefont {Dore}}, \bibinfo {author}
  {\bibfnamefont {J.}~\bibnamefont {Noronha}}, \ and\ \bibinfo {author}
  {\bibfnamefont {J.}~\bibnamefont {Noronha-Hostler}},\ }\href {\doibase
  10.1103/PhysRevC.105.L061901} {\bibfield  {journal} {\bibinfo  {journal}
  {Phys. Rev. C}\ }\textbf {\bibinfo {volume} {105}},\ \bibinfo {pages}
  {L061901} (\bibinfo {year} {2022})},\ \Eprint
  {http://arxiv.org/abs/2103.15889} {arXiv:2103.15889 [nucl-th]} \BibitemShut
  {NoStop}%
\bibitem [{\citenamefont {Chiu}\ and\ \citenamefont
  {Shen}(2021)}]{Chiu:2021muk}%
  \BibitemOpen
  \bibfield  {author} {\bibinfo {author} {\bibfnamefont {C.}~\bibnamefont
  {Chiu}}\ and\ \bibinfo {author} {\bibfnamefont {C.}~\bibnamefont {Shen}},\
  }\href {\doibase 10.1103/PhysRevC.103.064901} {\bibfield  {journal} {\bibinfo
   {journal} {Phys. Rev. C}\ }\textbf {\bibinfo {volume} {103}},\ \bibinfo
  {pages} {064901} (\bibinfo {year} {2021})},\ \Eprint
  {http://arxiv.org/abs/2103.09848} {arXiv:2103.09848 [nucl-th]} \BibitemShut
  {NoStop}%
\bibitem [{\citenamefont {da~Silva}\ \emph {et~al.}(2023)\citenamefont
  {da~Silva}, \citenamefont {Chinellato}, \citenamefont {Giannini},
  \citenamefont {Ferreira}, \citenamefont {Denicol}, \citenamefont {Hippert},
  \citenamefont {Luzum}, \citenamefont {Noronha},\ and\ \citenamefont
  {Takahashi}}]{daSilva:2022xwu}%
  \BibitemOpen
  \bibfield  {author} {\bibinfo {author} {\bibfnamefont {T.~N.}\ \bibnamefont
  {da~Silva}}, \bibinfo {author} {\bibfnamefont {D.~D.}\ \bibnamefont
  {Chinellato}}, \bibinfo {author} {\bibfnamefont {A.~V.}\ \bibnamefont
  {Giannini}}, \bibinfo {author} {\bibfnamefont {M.~N.}\ \bibnamefont
  {Ferreira}}, \bibinfo {author} {\bibfnamefont {G.~S.}\ \bibnamefont
  {Denicol}}, \bibinfo {author} {\bibfnamefont {M.}~\bibnamefont {Hippert}},
  \bibinfo {author} {\bibfnamefont {M.}~\bibnamefont {Luzum}}, \bibinfo
  {author} {\bibfnamefont {J.}~\bibnamefont {Noronha}}, \ and\ \bibinfo
  {author} {\bibfnamefont {J.}~\bibnamefont {Takahashi}} (\bibinfo
  {collaboration} {ExTrEMe}),\ }\href {\doibase 10.1103/PhysRevC.107.044901}
  {\bibfield  {journal} {\bibinfo  {journal} {Phys. Rev. C}\ }\textbf {\bibinfo
  {volume} {107}},\ \bibinfo {pages} {044901} (\bibinfo {year} {2023})},\
  \Eprint {http://arxiv.org/abs/2211.10561} {arXiv:2211.10561 [nucl-th]}
  \BibitemShut {NoStop}%
\bibitem [{\citenamefont {Krupczak}\ \emph {et~al.}(2024)\citenamefont
  {Krupczak} \emph {et~al.}}]{ExTrEMe:2023nhy}%
  \BibitemOpen
  \bibfield  {author} {\bibinfo {author} {\bibfnamefont {R.}~\bibnamefont
  {Krupczak}} \emph {et~al.} (\bibinfo {collaboration} {ExTrEMe}),\ }\href
  {\doibase 10.1103/PhysRevC.109.034908} {\bibfield  {journal} {\bibinfo
  {journal} {Phys. Rev. C}\ }\textbf {\bibinfo {volume} {109}},\ \bibinfo
  {pages} {034908} (\bibinfo {year} {2024})},\ \Eprint
  {http://arxiv.org/abs/2311.02210} {arXiv:2311.02210 [nucl-th]} \BibitemShut
  {NoStop}%
\bibitem [{\citenamefont {Domingues}\ \emph {et~al.}(2024)\citenamefont
  {Domingues}, \citenamefont {Krupczak}, \citenamefont {Noronha}, \citenamefont
  {da~Silva}, \citenamefont {Paquet},\ and\ \citenamefont
  {Luzum}}]{Domingues:2024pom}%
  \BibitemOpen
  \bibfield  {author} {\bibinfo {author} {\bibfnamefont {T.~S.}\ \bibnamefont
  {Domingues}}, \bibinfo {author} {\bibfnamefont {R.}~\bibnamefont {Krupczak}},
  \bibinfo {author} {\bibfnamefont {J.}~\bibnamefont {Noronha}}, \bibinfo
  {author} {\bibfnamefont {T.~N.}\ \bibnamefont {da~Silva}}, \bibinfo {author}
  {\bibfnamefont {J.-F.}\ \bibnamefont {Paquet}}, \ and\ \bibinfo {author}
  {\bibfnamefont {M.}~\bibnamefont {Luzum}},\ }\href@noop {} {\  (\bibinfo
  {year} {2024})},\ \Eprint {http://arxiv.org/abs/2409.17127} {arXiv:2409.17127
  [nucl-th]} \BibitemShut {NoStop}%
\bibitem [{\citenamefont {Baier}\ \emph {et~al.}(2008)\citenamefont {Baier},
  \citenamefont {Romatschke}, \citenamefont {Son}, \citenamefont {Starinets},\
  and\ \citenamefont {Stephanov}}]{BRSSS}%
  \BibitemOpen
  \bibfield  {author} {\bibinfo {author} {\bibfnamefont {R.}~\bibnamefont
  {Baier}}, \bibinfo {author} {\bibfnamefont {P.}~\bibnamefont {Romatschke}},
  \bibinfo {author} {\bibfnamefont {D.~T.}\ \bibnamefont {Son}}, \bibinfo
  {author} {\bibfnamefont {A.~O.}\ \bibnamefont {Starinets}}, \ and\ \bibinfo
  {author} {\bibfnamefont {M.~A.}\ \bibnamefont {Stephanov}},\ }\href {\doibase
  10.1088/1126-6708/2008/04/100} {\bibfield  {journal} {\bibinfo  {journal}
  {JHEP}\ }\textbf {\bibinfo {volume} {04}},\ \bibinfo {pages} {100} (\bibinfo
  {year} {2008})},\ \Eprint {http://arxiv.org/abs/0712.2451} {arXiv:0712.2451
  [hep-th]} \BibitemShut {NoStop}%
\bibitem [{\citenamefont {Kovtun}\ \emph {et~al.}(2005)\citenamefont {Kovtun},
  \citenamefont {Son},\ and\ \citenamefont {Starinets}}]{Kovtun:2004de}%
  \BibitemOpen
  \bibfield  {author} {\bibinfo {author} {\bibfnamefont {P.}~\bibnamefont
  {Kovtun}}, \bibinfo {author} {\bibfnamefont {D.~T.}\ \bibnamefont {Son}}, \
  and\ \bibinfo {author} {\bibfnamefont {A.~O.}\ \bibnamefont {Starinets}},\
  }\href {\doibase 10.1103/PhysRevLett.94.111601} {\bibfield  {journal}
  {\bibinfo  {journal} {Phys. Rev. Lett.}\ }\textbf {\bibinfo {volume} {94}},\
  \bibinfo {pages} {111601} (\bibinfo {year} {2005})},\ \Eprint
  {http://arxiv.org/abs/hep-th/0405231} {arXiv:hep-th/0405231} \BibitemShut
  {NoStop}%
\bibitem [{\citenamefont {Bazavov}\ \emph {et~al.}(2014)\citenamefont {Bazavov}
  \emph {et~al.}}]{HotQCD:2014kol}%
  \BibitemOpen
  \bibfield  {author} {\bibinfo {author} {\bibfnamefont {A.}~\bibnamefont
  {Bazavov}} \emph {et~al.} (\bibinfo {collaboration} {HotQCD}),\ }\href
  {\doibase 10.1103/PhysRevD.90.094503} {\bibfield  {journal} {\bibinfo
  {journal} {Phys. Rev. D}\ }\textbf {\bibinfo {volume} {90}},\ \bibinfo
  {pages} {094503} (\bibinfo {year} {2014})},\ \Eprint
  {http://arxiv.org/abs/1407.6387} {arXiv:1407.6387 [hep-lat]} \BibitemShut
  {NoStop}%
\bibitem [{\citenamefont {Bjorken}(1983)}]{Bjorken:1982qr}%
  \BibitemOpen
  \bibfield  {author} {\bibinfo {author} {\bibfnamefont {J.~D.}\ \bibnamefont
  {Bjorken}},\ }\href {\doibase 10.1103/PhysRevD.27.140} {\bibfield  {journal}
  {\bibinfo  {journal} {Phys. Rev. D}\ }\textbf {\bibinfo {volume} {27}},\
  \bibinfo {pages} {140} (\bibinfo {year} {1983})}\BibitemShut {NoStop}%
\bibitem [{\citenamefont {Heller}\ and\ \citenamefont
  {Spalinski}(2015)}]{heller2015hydrodynamics}%
  \BibitemOpen
  \bibfield  {author} {\bibinfo {author} {\bibfnamefont {M.~P.}\ \bibnamefont
  {Heller}}\ and\ \bibinfo {author} {\bibfnamefont {M.}~\bibnamefont
  {Spalinski}},\ }\href {\doibase 10.1103/PhysRevLett.115.072501} {\bibfield
  {journal} {\bibinfo  {journal} {Phys. Rev. Lett.}\ }\textbf {\bibinfo
  {volume} {115}},\ \bibinfo {pages} {072501} (\bibinfo {year} {2015})},\
  \Eprint {http://arxiv.org/abs/1503.07514} {arXiv:1503.07514 [hep-th]}
  \BibitemShut {NoStop}%
\bibitem [{\citenamefont {Baier}\ \emph {et~al.}(2006)\citenamefont {Baier},
  \citenamefont {Romatschke},\ and\ \citenamefont {Wiedemann}}]{Baier:2006um}%
  \BibitemOpen
  \bibfield  {author} {\bibinfo {author} {\bibfnamefont {R.}~\bibnamefont
  {Baier}}, \bibinfo {author} {\bibfnamefont {P.}~\bibnamefont {Romatschke}}, \
  and\ \bibinfo {author} {\bibfnamefont {U.~A.}\ \bibnamefont {Wiedemann}},\
  }\href {\doibase 10.1103/PhysRevC.73.064903} {\bibfield  {journal} {\bibinfo
  {journal} {Phys. Rev. C}\ }\textbf {\bibinfo {volume} {73}},\ \bibinfo
  {pages} {064903} (\bibinfo {year} {2006})},\ \Eprint
  {http://arxiv.org/abs/hep-ph/0602249} {arXiv:hep-ph/0602249} \BibitemShut
  {NoStop}%
\bibitem [{\citenamefont {Chattopadhyay}\ and\ \citenamefont
  {Heinz}(2020)}]{Chattopadhyay:2019jqj}%
  \BibitemOpen
  \bibfield  {author} {\bibinfo {author} {\bibfnamefont {C.}~\bibnamefont
  {Chattopadhyay}}\ and\ \bibinfo {author} {\bibfnamefont {U.~W.}\ \bibnamefont
  {Heinz}},\ }\href {\doibase 10.1016/j.physletb.2019.135158} {\bibfield
  {journal} {\bibinfo  {journal} {Phys. Lett. B}\ }\textbf {\bibinfo {volume}
  {801}},\ \bibinfo {pages} {135158} (\bibinfo {year} {2020})},\ \Eprint
  {http://arxiv.org/abs/1911.07765} {arXiv:1911.07765 [nucl-th]} \BibitemShut
  {NoStop}%
\bibitem [{\citenamefont {Romatschke}(2017)}]{Romatschke:2016hle}%
  \BibitemOpen
  \bibfield  {author} {\bibinfo {author} {\bibfnamefont {P.}~\bibnamefont
  {Romatschke}},\ }\href {\doibase 10.1140/epjc/s10052-016-4567-x} {\bibfield
  {journal} {\bibinfo  {journal} {Eur. Phys. J. C}\ }\textbf {\bibinfo {volume}
  {77}},\ \bibinfo {pages} {21} (\bibinfo {year} {2017})},\ \Eprint
  {http://arxiv.org/abs/1609.02820} {arXiv:1609.02820 [nucl-th]} \BibitemShut
  {NoStop}%
\bibitem [{\citenamefont {Romatschke}(2018)}]{Romatschke:2017vte}%
  \BibitemOpen
  \bibfield  {author} {\bibinfo {author} {\bibfnamefont {P.}~\bibnamefont
  {Romatschke}},\ }\href {\doibase 10.1103/PhysRevLett.120.012301} {\bibfield
  {journal} {\bibinfo  {journal} {Phys. Rev. Lett.}\ }\textbf {\bibinfo
  {volume} {120}},\ \bibinfo {pages} {012301} (\bibinfo {year} {2018})},\
  \Eprint {http://arxiv.org/abs/1704.08699} {arXiv:1704.08699 [hep-th]}
  \BibitemShut {NoStop}%
\bibitem [{\citenamefont {Florkowski}\ \emph {et~al.}(2018)\citenamefont
  {Florkowski}, \citenamefont {Heller},\ and\ \citenamefont
  {Spalinski}}]{Florkowski:2017olj}%
  \BibitemOpen
  \bibfield  {author} {\bibinfo {author} {\bibfnamefont {W.}~\bibnamefont
  {Florkowski}}, \bibinfo {author} {\bibfnamefont {M.~P.}\ \bibnamefont
  {Heller}}, \ and\ \bibinfo {author} {\bibfnamefont {M.}~\bibnamefont
  {Spalinski}},\ }\href {\doibase 10.1088/1361-6633/aaa091} {\bibfield
  {journal} {\bibinfo  {journal} {Rept. Prog. Phys.}\ }\textbf {\bibinfo
  {volume} {81}},\ \bibinfo {pages} {046001} (\bibinfo {year} {2018})},\
  \Eprint {http://arxiv.org/abs/1707.02282} {arXiv:1707.02282 [hep-ph]}
  \BibitemShut {NoStop}%
\bibitem [{\citenamefont {Jankowski}\ and\ \citenamefont
  {Spali\'nski}(2023)}]{Jankowski:2023fdz}%
  \BibitemOpen
  \bibfield  {author} {\bibinfo {author} {\bibfnamefont {J.}~\bibnamefont
  {Jankowski}}\ and\ \bibinfo {author} {\bibfnamefont {M.}~\bibnamefont
  {Spali\'nski}},\ }\href {\doibase 10.1016/j.ppnp.2023.104048} {\bibfield
  {journal} {\bibinfo  {journal} {Prog. Part. Nucl. Phys.}\ }\textbf {\bibinfo
  {volume} {132}},\ \bibinfo {pages} {104048} (\bibinfo {year} {2023})},\
  \Eprint {http://arxiv.org/abs/2303.09414} {arXiv:2303.09414 [nucl-th]}
  \BibitemShut {NoStop}%
\bibitem [{\citenamefont {Adams}\ \emph {et~al.}(2004)\citenamefont {Adams}
  \emph {et~al.}}]{STAR:2004moz}%
  \BibitemOpen
  \bibfield  {author} {\bibinfo {author} {\bibfnamefont {J.}~\bibnamefont
  {Adams}} \emph {et~al.} (\bibinfo {collaboration} {STAR}),\ }\href {\doibase
  10.1103/PhysRevC.70.054907} {\bibfield  {journal} {\bibinfo  {journal} {Phys.
  Rev. C}\ }\textbf {\bibinfo {volume} {70}},\ \bibinfo {pages} {054907}
  (\bibinfo {year} {2004})},\ \Eprint {http://arxiv.org/abs/nucl-ex/0407003}
  {arXiv:nucl-ex/0407003} \BibitemShut {NoStop}%
\bibitem [{\citenamefont {Chatrchyan}\ \emph {et~al.}(2012)\citenamefont
  {Chatrchyan} \emph {et~al.}}]{CMS:2012krf}%
  \BibitemOpen
  \bibfield  {author} {\bibinfo {author} {\bibfnamefont {S.}~\bibnamefont
  {Chatrchyan}} \emph {et~al.} (\bibinfo {collaboration} {CMS}),\ }\href
  {\doibase 10.1103/PhysRevLett.109.152303} {\bibfield  {journal} {\bibinfo
  {journal} {Phys. Rev. Lett.}\ }\textbf {\bibinfo {volume} {109}},\ \bibinfo
  {pages} {152303} (\bibinfo {year} {2012})},\ \Eprint
  {http://arxiv.org/abs/1205.2488} {arXiv:1205.2488 [nucl-ex]} \BibitemShut
  {NoStop}%
\bibitem [{\citenamefont {Betz}\ \emph
  {et~al.}(2009{\natexlab{a}})\citenamefont {Betz}, \citenamefont {Henkel},\
  and\ \citenamefont {Rischke}}]{Betz:2008me}%
  \BibitemOpen
  \bibfield  {author} {\bibinfo {author} {\bibfnamefont {B.}~\bibnamefont
  {Betz}}, \bibinfo {author} {\bibfnamefont {D.}~\bibnamefont {Henkel}}, \ and\
  \bibinfo {author} {\bibfnamefont {D.~H.}\ \bibnamefont {Rischke}},\ }\href
  {\doibase 10.1016/j.ppnp.2008.12.018} {\bibfield  {journal} {\bibinfo
  {journal} {Prog. Part. Nucl. Phys.}\ }\textbf {\bibinfo {volume} {62}},\
  \bibinfo {pages} {556} (\bibinfo {year} {2009}{\natexlab{a}})},\ \Eprint
  {http://arxiv.org/abs/0812.1440} {arXiv:0812.1440 [nucl-th]} \BibitemShut
  {NoStop}%
\bibitem [{\citenamefont {Betz}\ \emph
  {et~al.}(2009{\natexlab{b}})\citenamefont {Betz}, \citenamefont {Henkel},\
  and\ \citenamefont {Rischke}}]{Betz:2009zz}%
  \BibitemOpen
  \bibfield  {author} {\bibinfo {author} {\bibfnamefont {B.}~\bibnamefont
  {Betz}}, \bibinfo {author} {\bibfnamefont {D.}~\bibnamefont {Henkel}}, \ and\
  \bibinfo {author} {\bibfnamefont {D.~H.}\ \bibnamefont {Rischke}},\ }\href
  {\doibase 10.1088/0954-3899/36/6/064029} {\bibfield  {journal} {\bibinfo
  {journal} {J. Phys. G}\ }\textbf {\bibinfo {volume} {36}},\ \bibinfo {pages}
  {064029} (\bibinfo {year} {2009}{\natexlab{b}})}\BibitemShut {NoStop}%
\bibitem [{\citenamefont {Denicol}\ \emph {et~al.}(2012)\citenamefont
  {Denicol}, \citenamefont {Niemi}, \citenamefont {Molnar},\ and\ \citenamefont
  {Rischke}}]{Denicol:2012cn}%
  \BibitemOpen
  \bibfield  {author} {\bibinfo {author} {\bibfnamefont {G.~S.}\ \bibnamefont
  {Denicol}}, \bibinfo {author} {\bibfnamefont {H.}~\bibnamefont {Niemi}},
  \bibinfo {author} {\bibfnamefont {E.}~\bibnamefont {Molnar}}, \ and\ \bibinfo
  {author} {\bibfnamefont {D.~H.}\ \bibnamefont {Rischke}},\ }\href {\doibase
  10.1103/PhysRevD.85.114047} {\bibfield  {journal} {\bibinfo  {journal} {Phys.
  Rev. D}\ }\textbf {\bibinfo {volume} {85}},\ \bibinfo {pages} {114047}
  (\bibinfo {year} {2012})},\ \bibinfo {note} {[Erratum: Phys.Rev.D 91, 039902
  (2015)]},\ \Eprint {http://arxiv.org/abs/1202.4551} {arXiv:1202.4551
  [nucl-th]} \BibitemShut {NoStop}%
\bibitem [{\citenamefont {Monnai}\ and\ \citenamefont
  {Hirano}(2010)}]{Monnai:2010qp}%
  \BibitemOpen
  \bibfield  {author} {\bibinfo {author} {\bibfnamefont {A.}~\bibnamefont
  {Monnai}}\ and\ \bibinfo {author} {\bibfnamefont {T.}~\bibnamefont
  {Hirano}},\ }\href {\doibase 10.1016/j.nuclphysa.2010.08.002} {\bibfield
  {journal} {\bibinfo  {journal} {Nucl. Phys. A}\ }\textbf {\bibinfo {volume}
  {847}},\ \bibinfo {pages} {283} (\bibinfo {year} {2010})},\ \Eprint
  {http://arxiv.org/abs/1003.3087} {arXiv:1003.3087 [nucl-th]} \BibitemShut
  {NoStop}%
\bibitem [{\citenamefont {Tsumura}\ \emph {et~al.}(2007)\citenamefont
  {Tsumura}, \citenamefont {Kunihiro},\ and\ \citenamefont
  {Ohnishi}}]{Tsumura:2006hnr}%
  \BibitemOpen
  \bibfield  {author} {\bibinfo {author} {\bibfnamefont {K.}~\bibnamefont
  {Tsumura}}, \bibinfo {author} {\bibfnamefont {T.}~\bibnamefont {Kunihiro}}, \
  and\ \bibinfo {author} {\bibfnamefont {K.}~\bibnamefont {Ohnishi}},\ }\href
  {\doibase 10.1016/j.physletb.2006.12.074} {\bibfield  {journal} {\bibinfo
  {journal} {Phys. Lett. B}\ }\textbf {\bibinfo {volume} {646}},\ \bibinfo
  {pages} {134} (\bibinfo {year} {2007})},\ \bibinfo {note} {[Erratum:
  Phys.Lett.B 656, 274 (2007)]},\ \Eprint {http://arxiv.org/abs/hep-ph/0609056}
  {arXiv:hep-ph/0609056} \BibitemShut {NoStop}%
\bibitem [{\citenamefont {Tsumura}\ and\ \citenamefont
  {Kunihiro}(2010)}]{Tsumura:2009vm}%
  \BibitemOpen
  \bibfield  {author} {\bibinfo {author} {\bibfnamefont {K.}~\bibnamefont
  {Tsumura}}\ and\ \bibinfo {author} {\bibfnamefont {T.}~\bibnamefont
  {Kunihiro}},\ }\href {\doibase 10.1016/j.physletb.2010.05.041} {\bibfield
  {journal} {\bibinfo  {journal} {Phys. Lett. B}\ }\textbf {\bibinfo {volume}
  {690}},\ \bibinfo {pages} {255} (\bibinfo {year} {2010})},\ \Eprint
  {http://arxiv.org/abs/0906.0079} {arXiv:0906.0079 [hep-ph]} \BibitemShut
  {NoStop}%
\bibitem [{\citenamefont {Tsumura}\ and\ \citenamefont
  {Kunihiro}(2013)}]{Tsumura:2012ss}%
  \BibitemOpen
  \bibfield  {author} {\bibinfo {author} {\bibfnamefont {K.}~\bibnamefont
  {Tsumura}}\ and\ \bibinfo {author} {\bibfnamefont {T.}~\bibnamefont
  {Kunihiro}},\ }\href {\doibase 10.1103/PhysRevE.87.053008} {\bibfield
  {journal} {\bibinfo  {journal} {Phys. Rev. E}\ }\textbf {\bibinfo {volume}
  {87}},\ \bibinfo {pages} {053008} (\bibinfo {year} {2013})},\ \Eprint
  {http://arxiv.org/abs/1206.3913} {arXiv:1206.3913 [physics.flu-dyn]}
  \BibitemShut {NoStop}%
\bibitem [{\citenamefont {Jaiswal}(2013)}]{Jaiswal:2013npa}%
  \BibitemOpen
  \bibfield  {author} {\bibinfo {author} {\bibfnamefont {A.}~\bibnamefont
  {Jaiswal}},\ }\href {\doibase 10.1103/PhysRevC.87.051901} {\bibfield
  {journal} {\bibinfo  {journal} {Phys. Rev. C}\ }\textbf {\bibinfo {volume}
  {87}},\ \bibinfo {pages} {051901} (\bibinfo {year} {2013})},\ \Eprint
  {http://arxiv.org/abs/1302.6311} {arXiv:1302.6311 [nucl-th]} \BibitemShut
  {NoStop}%
\bibitem [{\citenamefont {Heller}\ \emph {et~al.}(2018)\citenamefont {Heller},
  \citenamefont {Kurkela}, \citenamefont {Spali\'nski},\ and\ \citenamefont
  {Svensson}}]{Heller:2016rtz}%
  \BibitemOpen
  \bibfield  {author} {\bibinfo {author} {\bibfnamefont {M.~P.}\ \bibnamefont
  {Heller}}, \bibinfo {author} {\bibfnamefont {A.}~\bibnamefont {Kurkela}},
  \bibinfo {author} {\bibfnamefont {M.}~\bibnamefont {Spali\'nski}}, \ and\
  \bibinfo {author} {\bibfnamefont {V.}~\bibnamefont {Svensson}},\ }\href
  {\doibase 10.1103/PhysRevD.97.091503} {\bibfield  {journal} {\bibinfo
  {journal} {Phys. Rev. D}\ }\textbf {\bibinfo {volume} {97}},\ \bibinfo
  {pages} {091503} (\bibinfo {year} {2018})},\ \Eprint
  {http://arxiv.org/abs/1609.04803} {arXiv:1609.04803 [nucl-th]} \BibitemShut
  {NoStop}%
\bibitem [{\citenamefont {Blaizot}\ and\ \citenamefont
  {Yan}(2018)}]{Blaizot:2017ucy}%
  \BibitemOpen
  \bibfield  {author} {\bibinfo {author} {\bibfnamefont {J.-P.}\ \bibnamefont
  {Blaizot}}\ and\ \bibinfo {author} {\bibfnamefont {L.}~\bibnamefont {Yan}},\
  }\href {\doibase 10.1016/j.physletb.2018.02.058} {\bibfield  {journal}
  {\bibinfo  {journal} {Phys. Lett. B}\ }\textbf {\bibinfo {volume} {780}},\
  \bibinfo {pages} {283} (\bibinfo {year} {2018})},\ \Eprint
  {http://arxiv.org/abs/1712.03856} {arXiv:1712.03856 [nucl-th]} \BibitemShut
  {NoStop}%
\bibitem [{\citenamefont {Behtash}\ \emph {et~al.}(2018)\citenamefont
  {Behtash}, \citenamefont {Cruz-Camacho},\ and\ \citenamefont
  {Martinez}}]{Behtash:2017wqg}%
  \BibitemOpen
  \bibfield  {author} {\bibinfo {author} {\bibfnamefont {A.}~\bibnamefont
  {Behtash}}, \bibinfo {author} {\bibfnamefont {C.~N.}\ \bibnamefont
  {Cruz-Camacho}}, \ and\ \bibinfo {author} {\bibfnamefont {M.}~\bibnamefont
  {Martinez}},\ }\href {\doibase 10.1103/PhysRevD.97.044041} {\bibfield
  {journal} {\bibinfo  {journal} {Phys. Rev. D}\ }\textbf {\bibinfo {volume}
  {97}},\ \bibinfo {pages} {044041} (\bibinfo {year} {2018})},\ \Eprint
  {http://arxiv.org/abs/1711.01745} {arXiv:1711.01745 [hep-th]} \BibitemShut
  {NoStop}%
\bibitem [{\citenamefont {Heller}\ and\ \citenamefont
  {Svensson}(2018)}]{Heller:2018qvh}%
  \BibitemOpen
  \bibfield  {author} {\bibinfo {author} {\bibfnamefont {M.~P.}\ \bibnamefont
  {Heller}}\ and\ \bibinfo {author} {\bibfnamefont {V.}~\bibnamefont
  {Svensson}},\ }\href {\doibase 10.1103/PhysRevD.98.054016} {\bibfield
  {journal} {\bibinfo  {journal} {Phys. Rev. D}\ }\textbf {\bibinfo {volume}
  {98}},\ \bibinfo {pages} {054016} (\bibinfo {year} {2018})},\ \Eprint
  {http://arxiv.org/abs/1802.08225} {arXiv:1802.08225 [nucl-th]} \BibitemShut
  {NoStop}%
\bibitem [{\citenamefont {Strickland}(2018)}]{Strickland:2018ayk}%
  \BibitemOpen
  \bibfield  {author} {\bibinfo {author} {\bibfnamefont {M.}~\bibnamefont
  {Strickland}},\ }\href {\doibase 10.1007/JHEP12(2018)128} {\bibfield
  {journal} {\bibinfo  {journal} {JHEP}\ }\textbf {\bibinfo {volume} {12}},\
  \bibinfo {pages} {128} (\bibinfo {year} {2018})},\ \Eprint
  {http://arxiv.org/abs/1809.01200} {arXiv:1809.01200 [nucl-th]} \BibitemShut
  {NoStop}%
\bibitem [{\citenamefont {Denicol}\ and\ \citenamefont
  {Noronha}(2020)}]{Denicol:2019lio}%
  \BibitemOpen
  \bibfield  {author} {\bibinfo {author} {\bibfnamefont {G.~S.}\ \bibnamefont
  {Denicol}}\ and\ \bibinfo {author} {\bibfnamefont {J.}~\bibnamefont
  {Noronha}},\ }\href {\doibase 10.1103/PhysRevLett.124.152301} {\bibfield
  {journal} {\bibinfo  {journal} {Phys. Rev. Lett.}\ }\textbf {\bibinfo
  {volume} {124}},\ \bibinfo {pages} {152301} (\bibinfo {year} {2020})},\
  \Eprint {http://arxiv.org/abs/1908.09957} {arXiv:1908.09957 [nucl-th]}
  \BibitemShut {NoStop}%
\bibitem [{\citenamefont {Heller}\ \emph {et~al.}(2020)\citenamefont {Heller},
  \citenamefont {Jefferson}, \citenamefont {Spali\'nski},\ and\ \citenamefont
  {Svensson}}]{Heller:2020anv}%
  \BibitemOpen
  \bibfield  {author} {\bibinfo {author} {\bibfnamefont {M.~P.}\ \bibnamefont
  {Heller}}, \bibinfo {author} {\bibfnamefont {R.}~\bibnamefont {Jefferson}},
  \bibinfo {author} {\bibfnamefont {M.}~\bibnamefont {Spali\'nski}}, \ and\
  \bibinfo {author} {\bibfnamefont {V.}~\bibnamefont {Svensson}},\ }\href
  {\doibase 10.1103/PhysRevLett.125.132301} {\bibfield  {journal} {\bibinfo
  {journal} {Phys. Rev. Lett.}\ }\textbf {\bibinfo {volume} {125}},\ \bibinfo
  {pages} {132301} (\bibinfo {year} {2020})},\ \Eprint
  {http://arxiv.org/abs/2003.07368} {arXiv:2003.07368 [hep-th]} \BibitemShut
  {NoStop}%
\bibitem [{\citenamefont {Kamata}\ \emph {et~al.}(2020)\citenamefont {Kamata},
  \citenamefont {Martinez}, \citenamefont {Plaschke}, \citenamefont
  {Ochsenfeld},\ and\ \citenamefont {Schlichting}}]{Kamata:2020mka}%
  \BibitemOpen
  \bibfield  {author} {\bibinfo {author} {\bibfnamefont {S.}~\bibnamefont
  {Kamata}}, \bibinfo {author} {\bibfnamefont {M.}~\bibnamefont {Martinez}},
  \bibinfo {author} {\bibfnamefont {P.}~\bibnamefont {Plaschke}}, \bibinfo
  {author} {\bibfnamefont {S.}~\bibnamefont {Ochsenfeld}}, \ and\ \bibinfo
  {author} {\bibfnamefont {S.}~\bibnamefont {Schlichting}},\ }\href {\doibase
  10.1103/PhysRevD.102.056003} {\bibfield  {journal} {\bibinfo  {journal}
  {Phys. Rev. D}\ }\textbf {\bibinfo {volume} {102}},\ \bibinfo {pages}
  {056003} (\bibinfo {year} {2020})},\ \Eprint
  {http://arxiv.org/abs/2004.06751} {arXiv:2004.06751 [hep-ph]} \BibitemShut
  {NoStop}%
\bibitem [{\citenamefont {Almaalol}\ \emph {et~al.}(2020)\citenamefont
  {Almaalol}, \citenamefont {Kurkela},\ and\ \citenamefont
  {Strickland}}]{Almaalol:2020rnu}%
  \BibitemOpen
  \bibfield  {author} {\bibinfo {author} {\bibfnamefont {D.}~\bibnamefont
  {Almaalol}}, \bibinfo {author} {\bibfnamefont {A.}~\bibnamefont {Kurkela}}, \
  and\ \bibinfo {author} {\bibfnamefont {M.}~\bibnamefont {Strickland}},\
  }\href {\doibase 10.1103/PhysRevLett.125.122302} {\bibfield  {journal}
  {\bibinfo  {journal} {Phys. Rev. Lett.}\ }\textbf {\bibinfo {volume} {125}},\
  \bibinfo {pages} {122302} (\bibinfo {year} {2020})},\ \Eprint
  {http://arxiv.org/abs/2004.05195} {arXiv:2004.05195 [hep-ph]} \BibitemShut
  {NoStop}%
\bibitem [{\citenamefont {Chattopadhyay}\ \emph {et~al.}(2022)\citenamefont
  {Chattopadhyay}, \citenamefont {Jaiswal}, \citenamefont {Du}, \citenamefont
  {Heinz},\ and\ \citenamefont {Pal}}]{Chattopadhyay:2021ive}%
  \BibitemOpen
  \bibfield  {author} {\bibinfo {author} {\bibfnamefont {C.}~\bibnamefont
  {Chattopadhyay}}, \bibinfo {author} {\bibfnamefont {S.}~\bibnamefont
  {Jaiswal}}, \bibinfo {author} {\bibfnamefont {L.}~\bibnamefont {Du}},
  \bibinfo {author} {\bibfnamefont {U.}~\bibnamefont {Heinz}}, \ and\ \bibinfo
  {author} {\bibfnamefont {S.}~\bibnamefont {Pal}},\ }\href {\doibase
  10.1016/j.physletb.2021.136820} {\bibfield  {journal} {\bibinfo  {journal}
  {Phys. Lett. B}\ }\textbf {\bibinfo {volume} {824}},\ \bibinfo {pages}
  {136820} (\bibinfo {year} {2022})},\ \Eprint
  {http://arxiv.org/abs/2107.05500} {arXiv:2107.05500 [nucl-th]} \BibitemShut
  {NoStop}%
\bibitem [{\citenamefont {Jaiswal}\ \emph {et~al.}(2022)\citenamefont
  {Jaiswal}, \citenamefont {Blaizot}, \citenamefont {Bhalerao}, \citenamefont
  {Chen}, \citenamefont {Jaiswal},\ and\ \citenamefont
  {Yan}}]{Jaiswal:2022udf}%
  \BibitemOpen
  \bibfield  {author} {\bibinfo {author} {\bibfnamefont {S.}~\bibnamefont
  {Jaiswal}}, \bibinfo {author} {\bibfnamefont {J.-P.}\ \bibnamefont
  {Blaizot}}, \bibinfo {author} {\bibfnamefont {R.~S.}\ \bibnamefont
  {Bhalerao}}, \bibinfo {author} {\bibfnamefont {Z.}~\bibnamefont {Chen}},
  \bibinfo {author} {\bibfnamefont {A.}~\bibnamefont {Jaiswal}}, \ and\
  \bibinfo {author} {\bibfnamefont {L.}~\bibnamefont {Yan}},\ }\href {\doibase
  10.1103/PhysRevC.106.044912} {\bibfield  {journal} {\bibinfo  {journal}
  {Phys. Rev. C}\ }\textbf {\bibinfo {volume} {106}},\ \bibinfo {pages}
  {044912} (\bibinfo {year} {2022})},\ \Eprint
  {http://arxiv.org/abs/2208.02750} {arXiv:2208.02750 [nucl-th]} \BibitemShut
  {NoStop}%
\bibitem [{\citenamefont {Brewer}\ \emph {et~al.}(2024)\citenamefont {Brewer},
  \citenamefont {Ke}, \citenamefont {Yan},\ and\ \citenamefont
  {Yin}}]{Brewer:2022ifw}%
  \BibitemOpen
  \bibfield  {author} {\bibinfo {author} {\bibfnamefont {J.}~\bibnamefont
  {Brewer}}, \bibinfo {author} {\bibfnamefont {W.}~\bibnamefont {Ke}}, \bibinfo
  {author} {\bibfnamefont {L.}~\bibnamefont {Yan}}, \ and\ \bibinfo {author}
  {\bibfnamefont {Y.}~\bibnamefont {Yin}},\ }\href {\doibase
  10.1103/PhysRevD.109.L091504} {\bibfield  {journal} {\bibinfo  {journal}
  {Phys. Rev. D}\ }\textbf {\bibinfo {volume} {109}},\ \bibinfo {pages}
  {L091504} (\bibinfo {year} {2024})},\ \Eprint
  {http://arxiv.org/abs/2212.00820} {arXiv:2212.00820 [nucl-th]} \BibitemShut
  {NoStop}%
\bibitem [{\citenamefont {Lifshitz}\ and\ \citenamefont
  {Pitaevskii}(1981)}]{LandauLifshitzPhysicalKinetics}%
  \BibitemOpen
  \bibfield  {author} {\bibinfo {author} {\bibfnamefont {E.~M.}\ \bibnamefont
  {Lifshitz}}\ and\ \bibinfo {author} {\bibfnamefont {L.~P.}\ \bibnamefont
  {Pitaevskii}},\ }\href@noop {} {\emph {\bibinfo {title} {Physical
  Kinetics}}},\ \bibinfo {edition} {2nd}\ ed.,\ \bibinfo {series} {Course of
  Theoretical Physics}, Vol.~\bibinfo {volume} {10}\ (\bibinfo  {publisher}
  {Butterworth-Heinemann},\ \bibinfo {year} {1981})\BibitemShut {NoStop}%
\bibitem [{\citenamefont {Werner}(2007)}]{Werner:2007bf}%
  \BibitemOpen
  \bibfield  {author} {\bibinfo {author} {\bibfnamefont {K.}~\bibnamefont
  {Werner}},\ }\href {\doibase 10.1103/PhysRevLett.98.152301} {\bibfield
  {journal} {\bibinfo  {journal} {Phys. Rev. Lett.}\ }\textbf {\bibinfo
  {volume} {98}},\ \bibinfo {pages} {152301} (\bibinfo {year} {2007})},\
  \Eprint {http://arxiv.org/abs/0704.1270} {arXiv:0704.1270 [nucl-th]}
  \BibitemShut {NoStop}%
\bibitem [{\citenamefont {Pierog}\ \emph {et~al.}(2015)\citenamefont {Pierog},
  \citenamefont {Karpenko}, \citenamefont {Katzy}, \citenamefont {Yatsenko},\
  and\ \citenamefont {Werner}}]{Pierog:2013ria}%
  \BibitemOpen
  \bibfield  {author} {\bibinfo {author} {\bibfnamefont {T.}~\bibnamefont
  {Pierog}}, \bibinfo {author} {\bibfnamefont {I.}~\bibnamefont {Karpenko}},
  \bibinfo {author} {\bibfnamefont {J.~M.}\ \bibnamefont {Katzy}}, \bibinfo
  {author} {\bibfnamefont {E.}~\bibnamefont {Yatsenko}}, \ and\ \bibinfo
  {author} {\bibfnamefont {K.}~\bibnamefont {Werner}},\ }\href {\doibase
  10.1103/PhysRevC.92.034906} {\bibfield  {journal} {\bibinfo  {journal} {Phys.
  Rev. C}\ }\textbf {\bibinfo {volume} {92}},\ \bibinfo {pages} {034906}
  (\bibinfo {year} {2015})},\ \Eprint {http://arxiv.org/abs/1306.0121}
  {arXiv:1306.0121 [hep-ph]} \BibitemShut {NoStop}%
\bibitem [{\citenamefont {Werner}\ \emph {et~al.}(2014)\citenamefont {Werner},
  \citenamefont {Guiot}, \citenamefont {Karpenko},\ and\ \citenamefont
  {Pierog}}]{Werner:2013tya}%
  \BibitemOpen
  \bibfield  {author} {\bibinfo {author} {\bibfnamefont {K.}~\bibnamefont
  {Werner}}, \bibinfo {author} {\bibfnamefont {B.}~\bibnamefont {Guiot}},
  \bibinfo {author} {\bibfnamefont {I.}~\bibnamefont {Karpenko}}, \ and\
  \bibinfo {author} {\bibfnamefont {T.}~\bibnamefont {Pierog}},\ }\href
  {\doibase 10.1103/PhysRevC.89.064903} {\bibfield  {journal} {\bibinfo
  {journal} {Phys. Rev. C}\ }\textbf {\bibinfo {volume} {89}},\ \bibinfo
  {pages} {064903} (\bibinfo {year} {2014})},\ \Eprint
  {http://arxiv.org/abs/1312.1233} {arXiv:1312.1233 [nucl-th]} \BibitemShut
  {NoStop}%
\bibitem [{\citenamefont {Werner}(2024)}]{Werner:2023jps}%
  \BibitemOpen
  \bibfield  {author} {\bibinfo {author} {\bibfnamefont {K.}~\bibnamefont
  {Werner}},\ }\href {\doibase 10.1103/PhysRevC.109.014910} {\bibfield
  {journal} {\bibinfo  {journal} {Phys. Rev. C}\ }\textbf {\bibinfo {volume}
  {109}},\ \bibinfo {pages} {014910} (\bibinfo {year} {2024})},\ \Eprint
  {http://arxiv.org/abs/2306.10277} {arXiv:2306.10277 [hep-ph]} \BibitemShut
  {NoStop}%
\bibitem [{\citenamefont {Kanakubo}\ \emph {et~al.}(2018)\citenamefont
  {Kanakubo}, \citenamefont {Okai}, \citenamefont {Tachibana},\ and\
  \citenamefont {Hirano}}]{Kanakubo:2018vkl}%
  \BibitemOpen
  \bibfield  {author} {\bibinfo {author} {\bibfnamefont {Y.}~\bibnamefont
  {Kanakubo}}, \bibinfo {author} {\bibfnamefont {M.}~\bibnamefont {Okai}},
  \bibinfo {author} {\bibfnamefont {Y.}~\bibnamefont {Tachibana}}, \ and\
  \bibinfo {author} {\bibfnamefont {T.}~\bibnamefont {Hirano}},\ }\href
  {\doibase 10.1093/ptep/pty129} {\bibfield  {journal} {\bibinfo  {journal}
  {PTEP}\ }\textbf {\bibinfo {volume} {2018}},\ \bibinfo {pages} {121D01}
  (\bibinfo {year} {2018})},\ \Eprint {http://arxiv.org/abs/1806.10329}
  {arXiv:1806.10329 [nucl-th]} \BibitemShut {NoStop}%
\bibitem [{\citenamefont {Kanakubo}\ \emph {et~al.}(2020)\citenamefont
  {Kanakubo}, \citenamefont {Tachibana},\ and\ \citenamefont
  {Hirano}}]{Kanakubo:2019ogh}%
  \BibitemOpen
  \bibfield  {author} {\bibinfo {author} {\bibfnamefont {Y.}~\bibnamefont
  {Kanakubo}}, \bibinfo {author} {\bibfnamefont {Y.}~\bibnamefont {Tachibana}},
  \ and\ \bibinfo {author} {\bibfnamefont {T.}~\bibnamefont {Hirano}},\ }\href
  {\doibase 10.1103/PhysRevC.101.024912} {\bibfield  {journal} {\bibinfo
  {journal} {Phys. Rev. C}\ }\textbf {\bibinfo {volume} {101}},\ \bibinfo
  {pages} {024912} (\bibinfo {year} {2020})},\ \Eprint
  {http://arxiv.org/abs/1910.10556} {arXiv:1910.10556 [nucl-th]} \BibitemShut
  {NoStop}%
\bibitem [{\citenamefont {Kanakubo}\ \emph
  {et~al.}(2022{\natexlab{a}})\citenamefont {Kanakubo}, \citenamefont
  {Tachibana},\ and\ \citenamefont {Hirano}}]{Kanakubo:2021qcw}%
  \BibitemOpen
  \bibfield  {author} {\bibinfo {author} {\bibfnamefont {Y.}~\bibnamefont
  {Kanakubo}}, \bibinfo {author} {\bibfnamefont {Y.}~\bibnamefont {Tachibana}},
  \ and\ \bibinfo {author} {\bibfnamefont {T.}~\bibnamefont {Hirano}},\ }\href
  {\doibase 10.1103/PhysRevC.105.024905} {\bibfield  {journal} {\bibinfo
  {journal} {Phys. Rev. C}\ }\textbf {\bibinfo {volume} {105}},\ \bibinfo
  {pages} {024905} (\bibinfo {year} {2022}{\natexlab{a}})},\ \Eprint
  {http://arxiv.org/abs/2108.07943} {arXiv:2108.07943 [nucl-th]} \BibitemShut
  {NoStop}%
\bibitem [{\citenamefont {Kanakubo}\ \emph
  {et~al.}(2022{\natexlab{b}})\citenamefont {Kanakubo}, \citenamefont
  {Tachibana},\ and\ \citenamefont {Hirano}}]{Kanakubo:2022ual}%
  \BibitemOpen
  \bibfield  {author} {\bibinfo {author} {\bibfnamefont {Y.}~\bibnamefont
  {Kanakubo}}, \bibinfo {author} {\bibfnamefont {Y.}~\bibnamefont {Tachibana}},
  \ and\ \bibinfo {author} {\bibfnamefont {T.}~\bibnamefont {Hirano}},\ }\href
  {\doibase 10.1103/PhysRevC.106.054908} {\bibfield  {journal} {\bibinfo
  {journal} {Phys. Rev. C}\ }\textbf {\bibinfo {volume} {106}},\ \bibinfo
  {pages} {054908} (\bibinfo {year} {2022}{\natexlab{b}})},\ \Eprint
  {http://arxiv.org/abs/2207.13966} {arXiv:2207.13966 [nucl-th]} \BibitemShut
  {NoStop}%
\bibitem [{\citenamefont {Gavassino}(2022)}]{Gavassino:2021owo}%
  \BibitemOpen
  \bibfield  {author} {\bibinfo {author} {\bibfnamefont {L.}~\bibnamefont
  {Gavassino}},\ }\href {\doibase 10.1103/PhysRevX.12.041001} {\bibfield
  {journal} {\bibinfo  {journal} {Phys. Rev. X}\ }\textbf {\bibinfo {volume}
  {12}},\ \bibinfo {pages} {041001} (\bibinfo {year} {2022})},\ \Eprint
  {http://arxiv.org/abs/2111.05254} {arXiv:2111.05254 [gr-qc]} \BibitemShut
  {NoStop}%
\bibitem [{\citenamefont {Murase}(2015)}]{Murase:2015oie}%
  \BibitemOpen
  \bibfield  {author} {\bibinfo {author} {\bibfnamefont {K.}~\bibnamefont
  {Murase}},\ }\emph {\bibinfo {title} {{Causal hydrodynamic fluctuations and
  their effects on high-energy nuclear collisions}}},\ \href {\doibase
  10.15083/00072981} {Ph.D. thesis},\ \bibinfo  {school} {Tokyo U.} (\bibinfo
  {year} {2015})\BibitemShut {NoStop}%
\bibitem [{\citenamefont {Sakai}\ \emph {et~al.}(2020)\citenamefont {Sakai},
  \citenamefont {Murase},\ and\ \citenamefont {Hirano}}]{Sakai:2020pjw}%
  \BibitemOpen
  \bibfield  {author} {\bibinfo {author} {\bibfnamefont {A.}~\bibnamefont
  {Sakai}}, \bibinfo {author} {\bibfnamefont {K.}~\bibnamefont {Murase}}, \
  and\ \bibinfo {author} {\bibfnamefont {T.}~\bibnamefont {Hirano}},\ }\href
  {\doibase 10.1103/PhysRevC.102.064903} {\bibfield  {journal} {\bibinfo
  {journal} {Phys. Rev. C}\ }\textbf {\bibinfo {volume} {102}},\ \bibinfo
  {pages} {064903} (\bibinfo {year} {2020})},\ \Eprint
  {http://arxiv.org/abs/2003.13496} {arXiv:2003.13496 [nucl-th]} \BibitemShut
  {NoStop}%
\bibitem [{\citenamefont {Sakai}\ \emph {et~al.}(2022)\citenamefont {Sakai},
  \citenamefont {Murase},\ and\ \citenamefont {Hirano}}]{Sakai:2021rug}%
  \BibitemOpen
  \bibfield  {author} {\bibinfo {author} {\bibfnamefont {A.}~\bibnamefont
  {Sakai}}, \bibinfo {author} {\bibfnamefont {K.}~\bibnamefont {Murase}}, \
  and\ \bibinfo {author} {\bibfnamefont {T.}~\bibnamefont {Hirano}},\ }\href
  {\doibase 10.1016/j.physletb.2022.137053} {\bibfield  {journal} {\bibinfo
  {journal} {Phys. Lett. B}\ }\textbf {\bibinfo {volume} {829}},\ \bibinfo
  {pages} {137053} (\bibinfo {year} {2022})},\ \Eprint
  {http://arxiv.org/abs/2111.08963} {arXiv:2111.08963 [nucl-th]} \BibitemShut
  {NoStop}%
\bibitem [{\citenamefont {Kuroki}\ \emph {et~al.}(2023)\citenamefont {Kuroki},
  \citenamefont {Sakai}, \citenamefont {Murase},\ and\ \citenamefont
  {Hirano}}]{Kuroki:2023ebq}%
  \BibitemOpen
  \bibfield  {author} {\bibinfo {author} {\bibfnamefont {K.}~\bibnamefont
  {Kuroki}}, \bibinfo {author} {\bibfnamefont {A.}~\bibnamefont {Sakai}},
  \bibinfo {author} {\bibfnamefont {K.}~\bibnamefont {Murase}}, \ and\ \bibinfo
  {author} {\bibfnamefont {T.}~\bibnamefont {Hirano}},\ }\href {\doibase
  10.1016/j.physletb.2023.137958} {\bibfield  {journal} {\bibinfo  {journal}
  {Phys. Lett. B}\ }\textbf {\bibinfo {volume} {842}},\ \bibinfo {pages}
  {137958} (\bibinfo {year} {2023})},\ \Eprint
  {http://arxiv.org/abs/2305.01977} {arXiv:2305.01977 [nucl-th]} \BibitemShut
  {NoStop}%
\end{thebibliography}%

\end{document}